\documentclass[a4paper,oneside,openright]{article}
\usepackage{epsf}
 \usepackage{epsfig}
\usepackage{srcltx}
\usepackage{jcappub} 
\usepackage{mathrsfs}
\usepackage{amsmath}
\usepackage{amssymb}
\usepackage{enumerate}
\usepackage{lipsum}
\usepackage{graphicx}
\usepackage{subfigure}
\usepackage{bm}
\usepackage{color}
\usepackage{setspace}
\usepackage{mathtools}
\usepackage{lscape}
\usepackage[11pt]{moresize}
\usepackage{afterpage}
\usepackage{lscape}
\usepackage{epsfig}
\usepackage{epstopdf}
\usepackage[titletoc,toc]{appendix}

\newcommand{\be}{\begin{equation}}
\newcommand{\ee}{\end{equation}} 
\newcommand{\eei}{\end{equation}\indent\indent}
\newcommand{\bc}{\begin{center}}
\newcommand{\ec}{\end{center}}
\newcommand{\ber}{\begin{eqnarray*}}
\newcommand{\ear}{\end{eqnarray*}}
\newcommand{\ba}{\begin{array}}
\newcommand{\ea}{\end{array}}
 
\newcommand{\na}{\nabla}

\newcommand{\ti}{\tilde}

\newcommand{\hatn}{a}

\newcommand{\bea}{\begin{eqnarray}}
\newcommand{\eea}{\end{eqnarray}}
\newcommand{\nn}{\nonumber}

\newcommand{\ei}{\end{itemize}}
\newcommand{\pd}{\partial}
\newcommand{\D}{{\mathrm{D}}}

\newcommand{\abs}[1]{\lvert#1\rvert}
\newcommand{\bra}[1]{\left(#1\right)}
\newcommand{\bras}[1]{\left[#1\right]}
\newcommand{\brac}[1]{\left\{#1\right\}}

\newcommand{\nab}{\nabla}
\newcommand{\la}{\langle}
\newcommand{\ra}{\rangle}

\newcommand \veps {\varepsilon} 

\newcommand{\MM}{{\cal M}}
\newcommand{\lb}{\{}
\newcommand{\rb}{\}}
\newcommand{\A}{{\cal A}}
\newcommand{\E}{{\cal E}}
\renewcommand{\H}{{\cal H}}
\newcommand{\R}{{\cal R}}
\newcommand{\f}{{\cal C}}

\newcommand{\Lietwo}{{\cal L}}

\newcommand{\reff}[1]{(\ref{#1})}

\renewcommand{\S}{_{\mathsf{S}}}
\newcommand{\V}{_{\mathsf{V}}}
\newcommand{\T}{_{\mathsf{T}}}
\newcommand{\vo}{{\mathbf{V_{\mathsf{O}}}}}
\newcommand{\ve}{{\mathbf{V_{\mathsf{E}}}}}

\newcommand*\xbar[1]{%
  \hbox{%
    \vbox{%
      \hrule height 0.5pt 
      \kern0.4ex
      \hbox{%
        \kern-0.2em
        \ensuremath{#1}%
        \kern-0.2em
      }%
    }%
  }%
} 

\def\bal#1\eal{\begin{align}#1\end{align}}
\def\case#1/#2{\textstyle\frac{#1}{#2} }

\def\fp{f^{\prime}}
\def\fpp{f^{\prime \prime}}
\def\fppp{f^{\prime \prime \prime}}

 {\begin{list}{}%
         {\setlength{\leftmargin}{#1} %
         \setlength{\rightmargin}{#2}%
         }%
         \item[\textit{}]%
 }
 {\end{list}}

\title{\boldmath Vibrating Black Holes in f(R) gravity}
\author[a]{Anne Marie Nzioki,}
\author[a]{Rituparno Goswami,}
\author[b,c]{Peter K. S. Dunsby}

\affiliation[a]{Astrophysics \& Cosmology Research Unit, 
School of Mathematics Statistics and Computer Science,
University of KwaZulu-Natal, 
Private Bag X54001, Durban 4000, South Africa.}
\affiliation[b]{Astrophysics Cosmology \&  Gravity Centre and 
Department of Mathematics and Applied Mathematics, 
University of Cape Town, Rondebosch,
7701, South Africa.}
\affiliation[c]{South African Astronomical Observatory, 
Observatory, Cape Town, South Africa.}

\emailAdd{anne.nzioki@gmail.com}
\emailAdd{Goswami@ukzn.ac.za}
\emailAdd{peter.dunsby@uct.ac.za}

\abstract{
We consider general perturbations of a Schwarzschild black holes in the context of $f(R)$ gravity.
A reduced set of frame independent master
variables are determined, which obey two closed wave equations
- one for the transverse, trace-free (tensor) perturbations and the other for the additional scalar degree of
freedom which characterise fourth-order theories of gravity.
We show that for the tensor modes, the underlying dynamics in
$f(R)$ gravity is governed by a modified Regge-Wheeler tensor
which obeys the same Regge-Wheeler equation as in General
Relativity. We find that the possible sources of scalar quasinormal modes  
that follow from scalar perturbations for the lower multipoles
result from primordial black holes, while higher mass, stellar
black holes are associated with extremely high multipoles,
which can only be produced in the first stage of black hole
formation. Since scalar quasi-normal modes are short ranged, this
scenario makes their detection beyond the range of current
experiments.}

\begin{document}
\sloppy
\maketitle
\nopagebreak

\section{Introduction}
\label{BH}
Einstein's theory of General Relativity (GR) \cite{Einstein:1916} is widely accepted to be one 
of the most successful fundamental theories in modern physics. Despite it's success, corrections to 
GR have been introduced recently to accommodate recent observations from the number counts of clusters of galaxies 
\cite{Allen:2011}, measurements of type Ia supernovae  \cite{Perlmutter:1999} and the Cosmic Microwave 
Background (CMB) anisotropies {\cite{Dunkley:2011}, which together seem to indicate that the energy density budget 
of the Universe comprises $ 5 \% $ ordinary matter (baryons, radiation and neutrinos), while the rest, which does not interact 
electromagnetically, consists of $ 27\% $ dark matter and $ 68 \% $ Dark Energy (DE) \cite{Planck:2013}. 
If GR is the correct theory of the gravitational action then its application to cosmology should incorporate these observations. 
Consequently, the simplest best fit model to our Universe is the  Friedmann-Lema\^{i}tre-Robertson-Walker (FLRW) model,
which is dominated by cold dark matter (CDM) and DE in the form of an effective cosmological constant, whose nature is still to 
be understood and is required to explain the late-time accelerated expansion of the Universe.  

One of the main motivations for exploring possible alternative theories of gravity arises from the obscure nature of DE candidates.
One possibility is to conjecture that the apparent need for DE could simply be a consequence of the break down of 
Einstein's equations on astrophysical and cosmological scales. One theory of modified gravity that has recently attracted 
a considerable amount of attention is fourth order gravity (FOG), which admits cosmologies that accelerate at late times without the presence of DE 
\cite{Capozziello:2002, Carroll:2004, Nojiri:2003, Starobinsky:2007, Capozziello:2003, Goheer:2009} and can and can account for the rotation curves for spiral 
galaxies without the need for dark matter \cite{Capozziello:2006a} (see \cite{Nojiri:2007, Sotiriou:2008, DeFelice:2010, Nojiri:2011, Clifton:2012} for detailed reviews).

The tetrad description of spacetime includes the Newman-Penrose null tetrad method 
\cite{Newman:1962} and the {\it 1+3 covariant approach} developed by Ehlers and Ellis 
\cite{Ehlers:1961,Ellis:1971,Ellis:1998} which includes both a full and `semi-tetrad'  
approach. The latter formalism is based on a 1+3 threading of the spacetime manifold 
with respect to a timelike congruence in such a way that tensorial objects encoding the physics can be 
decomposed into their space and time parts, and has been a useful tool for understanding of many aspects 
of relativistic fluid flows in cosmology and relativistic astrophysics. In particular the 1+3 approach to cosmological perturbation 
theory, developed by Ellis, Bruni and Dunsby \cite{Ellis:1990, Bruni:1992, Dunsby:1992}, which built on early work by 
Hawking \cite{Hawking:1966}, Lyth and Mukherjee \cite{Lyth:1988} and Ellis and Bruni \cite{Ellis:1989}, 
employs kinematic and dynamical variables to describe scalar, vector and tensor perturbations which have both a clear physical and geometric meaning 
and remain valid in all coordinate systems. This approach has been used to tackle problems in linear and non-linear perturbation theory and has been 
particularly successful in describing the physics of the CMB \cite{Dunsby:1997b, Challinor:1998, Gebbie:1999, Maartens:1999}. More recently, linear perturbation theory has been developed for describing the cosmology of fourth order theories of gravity  (FOG) using the 1+3 covariant approach \cite{Carloni:2008, Ananda:2008, Ananda:2009,  Abdelwahab:2008, Abebe:2012}, providing important features that differentiate the structure growth scenarios in FOG from standard GR.

A natural extension to the 1+3 approach, suitable for problems which have spherical symmetry, including the Schwarzschild solution, 
Lema\^{i}tre-Tolman-Bondi (LTB) models and many classes of Bianchi models was developed by Clarkson 
and Barrett \cite{Clarkson:2003}. This approach involves a `semi-tetrad' where, in addition to the timelike vector field of the 1+3 approach, a spatial vector is introduced. 
In GR, this `1+1+2 formalism' has been applied to the study of perturbations of locally rotationally symmetric (LRS) spacetimes 
\cite{Betschart:2004, Clarkson:2003, Clarkson:2004, Burston:2006, Clarkson:2007, Burston:2008, Zibin:2008, Dunsby:2010, Goswami:2011,
Goswami:2012,Goswami:2013} and strong lensing studies \cite{deSwardt:2010}. It has also been introduced to describe the properties of LRS spacetimes 
in the context of $f(R)$ gravity \cite{Nzioki:2010, Nzioki:2011, Nzioki:2014}. 

In GR, linear perturbations of black holes was first considered by Chandrasekhar using the metric approach together 
with the Newman-Penrose (NP) formalism \cite{Chandrasekhar} and more recently using the  `semi-tetrad' 1+1+2 covariant 
formalism by Clarkson and Barrett \cite{Clarkson:2003}. In the metric approach, perturbations of the Schwarzschild spacetime geometry are described by two wave equations, i.e., the Regge\,-Wheeler equation for odd parity modes and the Zerilli equation in the even parity case. 
These wave equations are expressed as functions (and their derivatives) in the perturbed metric which are not gauge-invariant, as a general coordinate transformation would not preserve the form of the wave equation. 
Using the 1+1+2 covariant approach, Clarkson and Barrett \cite{Clarkson:2003} 
demonstrated that both the odd and even parity perturbations may be unified in a covariant wave equation equivalent to the Regge\,-Wheeler 
equation. This wave equation is characterised by a single a covariant, frame- and gauge-invariant, transverse-traceless tensor. These results were extended 
to include couplings (at second order) to a homogeneous magnetic field leading to an accompanying electromagnetic signal alongside the standard tensor (gravitational wave modes) \cite{Clarkson:2004}.

There have been a number of recent investigations of the properties of black holes in FOG theories including an extensive study of the 
Schwarzschild de Sitter black hole in \cite{Cognola:2005, delaCruz:2009}. Perturbations of Schwarzschild black holes in $f(R,G)$ gravity were considered 
in \cite{DeFelice:2011} and a stability analysis of the Schwarzschild black hole in the Einstein Frame was presented in \cite{Myung:2011}.

The aim of this paper is to apply the 1+1+2 approach to the analysis of general linear perturbation of a
Schwarzschild black hole in $f(R)$ gravity. We perform all our calculations in the Jordan Frame, where the 
dynamics of the extra gravitational degree of freedom inherent in FOG theories is determined by the trace of the 
effective Einstein equations, leading to a linearised scalar wave equation for the Ricci scalar. Gauge invariance is assured by 
constructing perturbation variables which satisfy the Stewart-Walker lemma \cite{Stewart:1974} and we adopt the standard 
linearisation procedure by dropping all terms which are second order or higher in these variables.  Harmonic functions can then 
be introduced in the background which results in two decoupled parities reflecting the invariance of the background spacetime under 
parity transformation. The introduction of harmonics reduces the problem of finding a solution to one of simply solving a system of linear equations algebraically. After introducing the harmonic functions, the main objective is to find a reduced set of master variables which obey a closed set of wave equations. 

The outline of this paper is as follows. In Section \ref{chp:fR} we introduce $f(R)$ theories of gravity and present the general equations 
for these theories. Then in Section \ref{chp:Covariant} we outline the 1+3 and field 1+1+2 covariant methods in $f(R)$ gravity which provide 
a covariant (gauge invariant) description of spacetime. In Section \ref{chp:BH} we present the vacuum field equations 
linearised around a Schwarzschild black hole background using the 1+1+2 formalism. We 
discuss the spherical and time harmonics, which, when applied to our system of equations 
allows us to write them as a set of ordinary differential equations (ODEs) for each mode. Closed covariant and gauge-invariant wave 
equations for scalar and tensor modes are given in Section \ref{sec:RW}. In the case of tensors, this 
is just the Regge-Wheeler equation for a master variable that describes the evolution of a gauge and frame 
invariant transverse-traceless (TT) tensor. 

We then investigate the stability of these black hole to generic perturbations. Like in GR, initial tensor 
perturbations of the black hole eventually decay exponentially (ringing) at frequencies 
that are characteristic of the black hole and independent of the source of the perturbation -  a feature  first discovered by 
Vishveshwara in 1970 \cite{Vishveshwara:1970}. These {\it quasinormal modes} satisfy boundary conditions for purely 
outgoing waves at infinity and purely ingoing waves at the black hole horizon. In addition to these tensor modes, we also 
determine the quasinormal modes which arise from the additional scalar degree of freedom and discuss whether it is possible 
to use them to constrain $f(R)$ gravity. In Section \ref{chp:Sol} focuses on the method of solution to the perturbation equations using 
matrix methods where we demonstrate the significance of the freedom of choice of 
frame basis. Finally in section \ref{chp:Con} we present or our conclusions.

Unless otherwise specified, geometric units ($8\pi G=c=1$) will be used throughout this paper. 
The symmetrisation and the anti-symmetrisation  over the indexes of a tensor $ T_{ab}$ 
are defined as
\begin{equation}
T_{(a b)}= \frac{1}{2}\left(T_{a b}+T_{b a}\right)\;,\qquad T_{[a b]}= \frac{1}{2}\left(T_{a b}-T_{b a}\right)\,.
\end{equation}
over the indexes of the tensor. The symbol $\nabla$ represents the usual covariant derivative and $\partial$ corresponds to partial differentiation.

\section{$ f(R) $ Gravity} 
\label{chp:fR}
One of the most widely studied modifications to General Relativity is $f(R)$ gravity which is derived 
from the following action: 
\be
{\cal S}= \frac12 \int dV \bras{\sqrt{-g}\,f(R)+ 2\,\Lietwo_{M}(g_{ab}, \psi) } ~,
\label{action}
\ee 
where $R$ is the Ricci scalar. This represents the simplest generalisation of the 
Einstein-Hilbert action. Demanding that the action \reff{action} 
be invariant under a particular choice of symmetry guarantees that the resulting field equations 
also respect 
that symmetry. That being the case, since the Lagrangian is a function $R$ only, and $R$ is 
a generally covariant and a locally Lorentz 
invariant scalar quantity, then the field equations that follow are generally covariant and 
Lorentz invariant \reff{action}. After variation with respect to 
the metric $g_{ab}$  are given by:
\bea
\delta{\cal S} &=& -\frac12 \int dV  \, \sqrt{-g} \bras{\frac12 f \, g_{ab} \, \delta g^{ab} 
-\fp \,\delta R +T^{M}_{ab} \, \delta g^{ab} } ~,
\eea
where $ ' $ denotes differentiation with respect to $R$, and $ T^{M}_{ab} $ is the matter 
{energy momentum tensor} (EMT) defined as 
\be
T^{M}_{ab}=- \frac{2}{\sqrt{-g}} \, \frac{\delta \Lietwo_{M}} {\delta g^{ab} } ~.
 \label{metricEMT}
\ee
Writing the Ricci scalar as $R= g^{ab}\,R_{ab}$ and assuming the connection is the 
Levi-Civita one, we can write
\be
 \fp  \,\delta R \simeq  \delta g^{ab}\bra{\fp \, R_{ab}
 + g_{ab} \, \Box  \fp- \na_{a}\na_{b} \fp}~,
\ee
where the $\simeq$ sign denotes equality up to surface terms and 
$\Box \equiv \na_{c}\na^{c}$. By demanding that the action be stationary, so that 
$\delta{\cal S} =0$ with respect to variations in the metric, one has finally 
\be
\label{field1} 
 \fp \bra{R_{ab}-\frac12 g_{ab} \, R}=  \frac12 g_{ab}\,(f-R \,  \fp) 
 +\na_{a}\na_{b}\fp- g_{ab} \, \Box  \fp + T^{M}_{ab}~.
\ee
It can be seen that for the special case $f = R$, the equations reduce to the 
standard Einstein field equations.

It is convenient to write \reff{field1} in the form of effective Einstein equations as
\be
\label{field2} 
G_{ab} = \bra{R_{ab}-\frac12 g_{ab} \, R} 
= \tilde{T}^{M}_{ab} + T^{R}_{ab} = T_{ab}~,
\ee
where we define $T_{ab}$ as the total EMT comprising
\be
\tilde{T}^{M}_{ab} = \frac{T^{M}_{ab}}{ \fp} 
\label{matterEMT} 
\ee 
and
\be
T^{R}_{ab} = \frac{1}{ \fp} \bras{\frac12g_{ab} \,(f-R\, \fp) 
+\na_{a}\na_{b} \fp- g_{ab}\,\Box \fp}~. 
\label{curvatureEMT}
\ee
The components of the $T_{ab}$ can be considered to represent two effective 
``fluids" \cite{Capozziello:2002, Capozziello:2003,Capozziello:2006a,Carloni:2007}: 
the {curvature ``fluid"} (associated with $ T^{R}_{ab} $) and the 
{effective matter ``fluid"} (associated with $ \tilde{T}^{M}_{ab} $). This allows us to 
adapt more easily techniques from the ``covariant approach" 
(see, \cite{Ellis:1989, Dunsby:1992, Dunsby:1997a, Ellis:1998, Clarkson:2003}), to study 
a wide range of problems in $ f(R) $ gravity that were originally devised for GR.

The field equations \reff{field2} are fourth order in derivatives of the metric, which 
can be seen from the existence of the $\na_{a}\na_{b} \fp$ term in 
\reff{curvatureEMT}. This result also follows directly from a ramification of 
Lovelock's theorem \cite{Lovelock:1971, Lovelock:1972} which requires, in a 
four-dimensional Riemannian manifold, that the construction of a metric theory of modified 
gravity admits higher than second order derivatives to the 
field equations. This feature is problematic in a Lagrangian 
based theory as it can lead to Ostrogradski instabilities \cite{Ostrogradsky:1850} 
in the solutions of the field equations. In $f(R)$ theories, however, these   
instabilities are absent \cite{Woodard:2007}, due to the existence of an 
equivalence with scalar-tensor theories. 

In order to help avoid confusion later, we point out that we use the superscripts 
$^{M}$ and $^{R}$ to denote quantities relating to the standard matter fluid and 
curvature fluid respectively and that the unbarred dynamic quantities with no 
superscripts are derived from the total effective EMT.

\section{Formalism}
\label{chp:Covariant}
\subsection{The 1+3 formalism}
The covariant approach we consider adopts a fluid-flow description of the matter content (including any modifications to General Relativity)
of spacetime. In the usual 1+3 splitting \cite{Ehlers:1961, Kundt:1962, Ellis:1971, Ellis:1973, Maartens:1997a} 
of spacetime,  the fluid flow is determined at each point by the field  vector $u^a$, tangent to the flow lines. The vector $u^a$ is a timelike unit vector representing
 the normalised 4-velocity of the matter, hence
\be
u^a\, u_a = -1~.
\ee
The tensor
\be
h_{ab} \equiv g_{ab} +u_a\,u_b ~, 
\ee
projects any tensor onto the hypersurface orthogonal to $u^a$ and has the following properties
\be
h_{ab}\, u^{b} = 0 ~,\quad  h_{a}{}^{c} \, h_{c}{}^{b} = h_{a}{}^{b}~,
\quad h_{a}{}^{a} = 3~.
\ee
These constant time hypersurfaces represent the 
local rest 3-space associated with the observer.

The effective {volume element} for the rest space of the comoving observer is given by
\be
\veps_{abc} =\veps_{abcd}\,  u^{d} ~, \qquad \mathrm{where} 
\qquad \veps_{abc}= \veps_{[abc]}   ~~ \mathrm{and} ~~  
 \veps_{abc}\, u^{c}= 0 ~,
 \label{volume1}
\ee
where $ \veps_{abcd} $ is the four-dimensional volume element $ (\veps_{abcd}= \sqrt{\mid \mathrm{det}~ g\mid} )
\delta^{0}{}_{[a} \,\delta^{1}{}_{b}  \,\delta^{2}{}_{c}  \,\delta^{3}{}_{d]} )$ of the spacetime manifold. 

Any projected rank-2 tensor $ S_{ab}$ can be split as 
\be
S_{ab} = S_{\la ab \ra}  + \frac13 S\,  h_{ab}  + S_{[ ab ]}~, 
\label{three_tensor}
\ee
where $ S = {h}_{ab}  S^{ab} $ is the spatial trace, $S_{\la ab \ra}$ is the orthogonally 
{projected symmetric trace-free} PSTF part of the tensor defined as
\be
S_{\la ab \ra} = \bra{h_{c} {}^{(a}\,  h_{d}{}^{b)}- \frac{1}{3} h_{ab}\, h^{cd}} S_{cd} ~,
\label{PSTF}
\ee
and $S_{[ ab ]}$ is the antisymmetric part of this tensor.  We use angle brackets to represent any PSTF tensors.

The covariant derivatives for any tensor $ S_{a..b}{}^{c..d} $ are defined as the time derivative 
along $u^a$:
\be
\label{dot} 
\dot{S}_{a..b}{}^{c..d} \equiv  u^{f}\,\nab_{f} S_{a..b}{}^{c..d}~, 
\ee 
and the covariant spatial derivative defined in the local rest 3-spaces orthogonal to $u^a$:
\be
\label{spatial} 
\D_{e}S_{a..b}{}^{c..d} = h_{e}{}^{j} \,h_{a}{}^{l}\,...\,h_{b}{}^g\,h_{f}{}^{c}\,\,...\,
h_i{}^d\,\D_j S_{l..g}{}^{f..i}~, 
\ee
with projection on all the free indices. 

Kinematical quantities are introduced by decomposing the covariant derivative of 
$u^a$ into its irreducible parts:
\be
\label{velocitygrad}
\na_{a} u_{b} = - u_{a}\,  \dot{u}_{b}+	\sigma_{ab} + \omega_{ab} + \frac{1}{3} \Theta\, h_{ab} ~. 
\ee
where
\bea
\dot{u}_{b} = u^{c}\, \nab_{c} u_{b} ~, \qquad
\Theta = \D^{a} u_{a}~,\qquad
\omega_{ab} = \D_{[a} u_{b]}~, \qquad
\sigma_{ab} = \D_{\la a} u_{b\ra}~,
\eea
are respectively, the four-acceleration, the expansion scalar which 
represents the local volume rate of expansion of the fluid,
the antisymmetric vorticity tensor which describes the rigid rotation of matter 
relative to a non-rotating frame, 
and the PSTF shear tensor that determines the distortion arising in the matter 
flow, leaving the volume invariant.
By construction, the following properties hold for these kinematical quantities
\bea
\sigma_{[ab]} = \omega_{[ab]} = 0~, \qquad \omega_{ab}\,  u^{b} =  \sigma_{ab} \, u^{b} = 0~,
 \qquad \sigma^{a} {}_{a} = 0 ~.
\eea
The total {energy momentum tensor} (EMT) $T_{ab}$ as defined in \reff{field2}
can be decomposed relative to $ u^{a}$ by splitting it into parts parallel and 
orthogonal to $ u^{a}$ as follows:
\be
\label{EMT1}
T_{ab} = \mu  \,u_{a}  \,u_{b} + q_{a}  \,u_{b} + u_{a} \, q_{b} + p  \,h_{ab} + \pi_{ab} ~;  \,
\ee
where $\mu$ is the total effective {energy density} relative to $ u^{a}$, $p$ the total 
{isotropic pressure}, $q_{a}$ the total energy flux (momentum density) relative to $ u^{a}$ 
and $\pi_{ab}$ the and PSTF total {anisotropic stress}, such that
\bea
\mu \quad= & T_{ab} \, u^{a} \, u^{b} &= \frac{\mu^{M}}{\fp} + \mu^{R}~,
 \qquad p\quad = \frac{1}{3}\, T_{ab} \, h^{ab} = \frac{p^{M}}{\fp} + p^{R}~,\\
q_{a} \quad=& \quad- \,T_{bc}  \,u^{c}  \,h^{b}{}_{a} &= \frac{q^{M}_{a}}{\fp} + q^{R}_{a}~,
 \qquad \pi_{ab}\quad = T_{cd}  \,h^{c} {}_{\la a} \, h^{d}{}_{b \ra} =  \frac{\pi^{M}_{ab}}{\fp} + \pi^{R}_{ab}~.
\eea
An equation of state needs to be specified to relate the matter thermodynamic variables.

The derivative terms of the curvature EMT $T^{R}_{ab}$  can be decomposed into time and 
spatial parts resulting in
\bea
\label{decomposedEMT}
T^{R}_{ab} &=& \frac{1}{\fp} \bras{\frac{1}{2}g_{ab} \,(f - R \, \fp) 
-  \dot{\fp} \bra{ \frac13 h_{ab} \, \theta + \sigma_{ab}  + \omega_{ab}}
 + \frac13 h_{ab} \,\D^{2} \fp     \right. \nn\\
 && \left. +\, \D_{\la a}  \D_{b\ra}  \fp+ \frac12 \veps_{abc} \,\veps^{cdf}  \D_{d} \D_{f} \fp
 - u_{a}  \bra{  h_{cb} \, (\D^{c} \fp)^{\dot{}}+ \dot{u}_{c}  \,u_{b}  \,\D^{c}  \fp
  - \dot{\fp} \,  \dot{u}_{b}} \right. \nn\\ 
 &&\left. + \,u_{b} \bra{ \frac13 \theta  \,\D_{a}  \fp+ \sigma_{a}{}^{c} \,\D_{c}  \fp 
 + \omega_{a}{}^{c} \, \D_{c}  \fp +u_{a} \, \ddot{\fp}  - \D_{a}  \dot{\fp}}   \right. \nn\\
&& \left.   -\, g_{ab}\bra{\dot{u}_{c}  \,\D^{c}  \fp  -  \theta  \, \dot{\fpp} - \ddot{\fp}  + \D^{2} \fp }}  ~. 
\eea
The locally free gravitational field is given by the {Weyl curvature tensor} $C_{abcd}$ 
defined by the equation
\be
C^{ab}{}_{cd} = R^{ab}{}_{cd}-2 g^{[a}{}_{[c}\, R^{b]}{}_{d]} + \frac13 R\, g^{[a}{}_{[c} \,g^{b]}{}_{d]} ~. 
\label{Weyl}
\ee
which can be split relative to $u^a$ into the `electric' and `magnetic' Weyl curvature parts as follows:
\be
E_{ab} = C_{abcd} ~, \qquad  
H^{a}{}_a = 0, ~,
\ee
in analogy to the 1+3 split of the Maxwell field strength tensor \cite{Maartens:1998}.

The dynamical relations for an arbitrary spacetime in the 1+3 formulation of FOG arise 
from the Ricci identities for the fundamental timelike vector field $u^{a}$, that is,
\be
2\,\na_{[a}\na_{b]}u^{c} = R_{ab}{}^{c}{}_{d}\,u^{d}~, \label{Ricci2}
\ee 
and from contracting the second Bianchi identities
\be
\label{Bianchi2}
 \nabla_{[e}\,R_{ab]cd}=0~. 
\ee
resulting in a set of propagation and constraint equations when covariantly decomposed 
\cite{vanElst:1996}. 
We have to include the trace of \reff{field1}
\be
\label{Curvetrace1}
R\,\fp-2f = - \,3 \bra{\fpp \,\D^{2}R + \fppp \,\D^{a}R\, \D_{a}R - \fppp\, \dot{R^{2}} - \fpp \,\ddot R 
+ \dot{u}_{c}\,\fpp \D^{c} R -  \fpp \,\theta\, \dot{R} }~.
\ee
in order to close the system of equations.

\subsection{The 1+1+2 formalism}
The 1+3 covariant approach involves splitting spacetime into its temporal and spatial parts
in such a way that the local 3-space is orthogonal to the vector field $u^a$ which provides a timelike 
threading for the spacetime. This can be naturally extended to give a 1+1+2 covariant 
decomposition of spacetime by introducing
the unit vector field $n^a$ in the local 3-space orthogonal to $u^a$, such that
\be
n^a\, n_a = 1, \quad n^a\, u_a = 0\;.
\ee
The 2-dimensional tensor
\be
N_{ab} \equiv g_{ab} +u_a\,u_b  - n_a\,n_b ~, 
\qquad n^a \,N_{ab} = 0 =u^{a} \, N_{ab} ~,
\qquad N^{a}{}_{a} = 2~.
\ee
projects onto the tangent 2-spaces (which we call `sheets') orthogonal to both $u^a$ and $n^a$.
The volume element of the sheet is the totally anti-symmetric 2-tensor  
\be
\veps_{ab}\equiv\veps_{abc}\, n^{c} ~, 
\label{perm}
\ee 
where $\veps_{abc}$ is the volume element of the 3-spaces.

The covariant derivatives for any tensor $ {S}_{a..b}{}^{c..d} $ are defined as the time derivative `$~^.~$' along $u^a$ as given in \reff{dot},  
the spatial divergence `$~\hat{}~$' along $n^a$ in the surfaces orthogonal to $u^a$
\be 
\hat{S}_{a..b}{}^{c..d} \equiv  n^{f}\,D_{f} S_{a..b}{}^{c..d}~,
\ee
and the projected covariant derivative `$\delta_a$' on the sheet
\be
\delta_f S_{a..b}{}^{c..d} \equiv  
N_{f}{}^{j} \,N_{a}{}^{l}\,...\,N_{b}{}^g\,N_{h}{}^{c}\,\,...\,
N_i{}^d\,\D_j S_{l..g}{}^{h..i}~,
\ee
where again the projection applies to every free index. The spatial derivative `$D_a$' is as 
defined in \reff{spatial}. 

In the 1+1+2 splitting of spacetime, any 3-vector $V^{a}$ can be irreducibly split into a scalar 
component, $\cal{V}$, along $n^{a}$ and a 2-vector component on the sheet, ${\cal{V}}^{a}$, 
orthogonal to $n^{a}$, i.e., 
\be
\label{3-vecsplit}  
V^{a} = {\cal{V}} \, n^{a} + {\cal{V}}^{a}~, \qquad \mathrm{where} \quad {\cal{V}}\equiv V_{a} \,n^{a}
~~{\mathrm{and}}~~{\cal{V}}^{a} \equiv N^{ab} \, V_{b}  ~, 
\ee
Similarly, a PSTF 3-tensor, $V_{ab}$, can be decomposed into 2-scalar, 
2-vector and PSTF 2-tensor parts as
\be 
\label{3-tensorsplit}
V_{ab} = V_{\la ab\ra} = {\cal{V}} \bra{n_{a} \,n_{b} - \frac12 N_{ab}}
+2 {\cal{V}}_{(a} \, n_{b)} + {\cal{V}}_{ab}~,    
\ee 
where 
\be
{\cal{V}} \equiv  n^{a}\,n^{b} \,V_{ab}= -N^{ab} V_{ab}~, \qquad
{\cal{V}}_{a} \equiv  N_{a}{}^{b}\, n^c \, V_{bc}~,\qquad
{\cal{V}}_{ab} \equiv  V_{\{ab\}} \equiv \bra{ N_{(a}{}^{c}\,N_{b)}{}^{d} 
- \frac{1}{2}N_{ab}\, N^{cd}} V_{cd}~.
\ee
The curly brackets denote the part of a tensor which is PSTF with respect to $n^{a}$. 

It then follows that the 1+3 kinematical and Weyl quantities can be irreducibly split as
\bea
\dot{u}^a &=&\A \, n^a+\A^a\ ,\\
\omega^a&=&\Omega \,n^a+\Omega^a\ ,\\
\sigma_{ab}&=&\Sigma \bra{n_a\,n_b-\frac12 N_{ab}} +2\,\Sigma_{(a}\,n_{b)}+\Sigma_{ab}\ ,\\
E_{ab}&=&{\E}\bra{n_a\,n_b-\frac12 N_{ab}} +2\,{\E}_{(a}\,n_{b)}+{\E}_{ab}\ ,\\
H_{ab}&=&{\H}\bra{n_a\,n_b-\frac12 N_{ab}}+2\,{\H}_{(a}\,n_{b)}+{\H}_{ab}\ .
\eea
The irreducible form of the covariant decomposition of the derivative of $n^a$ is
\be
\label{fullcov_n}
\na_a\,n_b=-\A\, u_a\,u_b-u_a\,\alpha_b + \bra{\frac13\Theta+\Sigma}n_{a}\,u_{b} 
+ \bra{\Sigma_a-\veps_{ac}\,\Omega^c}\,u_b 
+n_a\,\hatn_b +\frac12\phi\,N_{ab}+\xi\,\veps_{ab}+\zeta_{ab}\ , 
\ee
where along the spatial direction $n^{a}$, $ \phi =  \delta_{a}n^{a}$ is the expansion of the sheet, 
$\zeta_{ab} = \delta_{\lb a}n_{b \rb }$ is the shear of $n^{a}$
and $ \hatn_a = n^c\,\D_c\, n_a= \hat n_a$ its acceleration, while 
$\xi = \frac{1}{2}\, \veps^{ab}\delta_{a}n_{b}$ is the 
vorticity associated with $n^{a}$. 

Finally,  the anisotropic fluid variables $q_{a}$ and $\pi_{ab}$ can be split as follows:
\bea
\label{flux112}
q_{a} &=& Q \,n_{a} + Q_{a}~, \\
\label{aniso112}
\pi_{ab} &=& \Pi \bra{ n_{a}\,n_{b} -\frac12 N_{ab}} + 2\Pi_{(a}\,n_{b)} + \Pi_{ab}~.
\eea

\subsubsection{Energy momentum tensor}
In terms of the 1+1+2 variables, the total energy momentum tensor \reff{EMT1} is given by
\be
\label{EMTsplit} \,
T_{ab} = \mu \,u_{a} \,u_{b} + p\,h_{ab} 
+ 2 u_{(a}\bras{ Q\, n_{b)} + Q_{a}}  +  \Pi \bra{ n_{a}\,n_{b} 
-\frac12 N_{ab}} + 2\, \Pi_{(a}\,n_{b)} + \Pi_{ab}\ . 
\ee 
Moreover, in terms of the 1+1+2 variables, the curvature fluid can be decomposed as follows: 
\bea
 \label{generaldensity}
\mu^{R}  
&=& \frac{1}{\fp} \bras{\frac12 (R\, \fp-f) - \theta  \fpp \,\dot{R} 
+ \fppp \, X^{2} + \fppp \,\delta^{a} R \,\delta_{a} R + \, \fpp \, \hat{ X} + \phi \fpp \, X  
 - \hatn^{a} \,\fpp \, \delta_{a}R + \fpp \, \delta^{a}\delta_{a} R  } ~, 
 \eea
 \bea
\label{generalpressure}
p^{R} &=& \frac{1}{\fp} \bras{\frac12 (f-R\, \fp) + \frac23 \theta \, \fpp \dot{R}
+ \fppp \,\dot{R}^{2} + \fpp \,\ddot{R}  - \A \,\fpp\, X  -\,\A^{a} \fpp \, \delta_{a} R  -\frac23( \phi \, \fpp\, X 
+ \fppp \, \delta^{a} R\, \delta_{a} R  \right. \nn \\  
&&\left. + \, \fpp \, \delta^{a} \delta_{a} R +  \fppp \, X^{2} +  \fpp \,\hat{X} 
- \hatn_{a} \, \fpp \, \delta^{a}R)} ~,
\eea
 \bea
 \label{generalfluxSc}
Q^{R} 
&=&- \,\frac{1}{\fp} \bras{  \fppp \, \dot{R} \,X 
+ \fpp \, \bra{ \dot{X}-\A \,\dot{R} } -\alpha^{a}\,\fpp\, \delta_{a}R } ~,
\eea
 \bea
 \label{generalfluxVc}
Q_{a}^{R}
&=&	 \frac{1}{\fp} \bras{ \bra{ \frac13 \theta
-\frac12 \Sigma}  \fpp \, \delta_{a} R 
+\bra{ \Sigma_{a} -\veps_{a}{}^{b}  \Omega_{b}} \fpp \, X
 +\, \bra{ \Sigma_{a}{}^{b} + \veps_{a}{}^{b} \Omega}\, \fpp \, \delta_{b} R   
-\dot{R}\, \fppp \, \delta_{a}R - \fpp \, \delta_{a} \dot{R} }  ~,  
\eea
 \bea
 \label{generalstressSc}
\Pi^{R} 
&=& \frac{1}{\fp} \bras{ \frac13 \bra{ 2 \fppp \,X^{2} 
+2 \fpp \,\hat{X} - 2 \A_{a}\, \fpp \, \delta^{a} R - \phi \,\fpp \, X 
-\, \fppp \, \delta^{a}R \,\delta_{a} R  
- \fpp \, \delta^{a} \delta_{a} R } -  \Sigma \,  \fpp \, \dot{R} } ~,
\eea
 \bea
 \label{generalstressVc}
\Pi_{a}^{R} 
&=&  \frac{1}{\fp} \bras{ - \Sigma_{a}\, \fpp \, \dot{R} 
+X \, \fppp \, \delta_{a} R + \fpp \, \delta_{a} X 
- \frac12 \phi \, \fpp \, \delta_{a} R 
 +\,\bra{ \xi \,\veps_{a}{}^{b} - \zeta_{a}{}^{b} }\fpp \,\delta_{b} R 
- \frac12 \bra{ \Sigma_{a} 
+ \veps_{a}{}^{b} \Omega_{b}}\,\fpp \, \dot{R}  } ~,
\eea
 \bea
 \label{generalstressTn}
\Pi_{ab}^{R}
&=&\frac{1}{\fp} \bras{- \Sigma_{ab} \,  \fpp \, \dot{R} 
+ \zeta_{ab}\,  \fpp \, X  + \, \fppp \, \delta_{\lb a} R\, \delta_{b \rb} R  
+  \fpp \, \delta_{\lb a} \delta_{b \rb} R} ~,
\eea
where we have defined $\hat{R}=X$.
Additionally, the 1+1+2 split of the curvature trace equation
\reff{Curvetrace1} results in
\be
\label{Curvetrace2}
R\,\fp-2 f = 3 \bra{ \fpp \,\theta\, \dot{R} - \fppp \,X^{2} 
- \fppp \,\delta^{a}R\, \delta_{a}R 
 -(\A +\phi) \fpp \,X   -\fpp \, \hat{X} 
-   \fpp \, \delta^{a} \delta_{a} R 
+ \fppp\, \dot{R^{2}} + \fpp \,\ddot R  }~.
\ee
\subsubsection{Commutation relations}
The three derivatives defined so far, dot - $`\dot{\phantom{x}}$', 
hat - $`\hat{\phantom{x}}$' and delta - $`\delta_a$' satisfy the following commutation 
relations when they act on scalars ${\cal{V}}$:
\bea
\label{comm1}
\hat{\dot {\mathcal{V}}}-\dot{\hat {\cal{V}}} &=&-\A\,\dot{\cal{V}}+\bra{\frac13 \Theta 
+ \Sigma}\hat{\cal{V}}+ \bra{\Sigma_a + \veps_{ab}\Omega^b
-\alpha_a}\delta^a{\cal{V}} ~,
\\
\label{comm2}
\delta_a\dot{\cal{V}}-\bra{\delta_a{\cal{V}}}^{~\cdot}_{\perp} &=&
-\A_{a}\,\dot{\cal{V}} + \bra{\alpha_{a} +\Sigma_a 
- \veps_{ab}\Omega^b}\hat{\cal{V}} 
+ \bra{\frac13\Theta-\frac12\Sigma}\delta_a{\cal{V}} 
 + \bra{\Sigma_{ab}+\Omega\,\veps_{ab}}\delta^b{\cal{V}}\  ~, 
 \\
 \label{comm3}
\delta_a\hat{\cal{V}}-(\widehat{\delta_a{\cal{V}}})_{\perp}  &=& 
-2\,\veps_{ab}\Omega^b\,\dot{\cal{V}}+\hatn_a\,\hat{\cal{V}}
+\frac12\phi\,\delta_a{\cal{V}} +\bra{\zeta_{ab}+\xi\,\veps_{ab}}\delta^b{\cal{V}}\  ~,
  \\
  \label{comm4}
\delta_{[a}\delta_{b]}{\cal{V}} &=&\veps_{ab}\bra{\Omega\, \dot{\cal{V}}-\xi\,\hat{\cal{V}}}~.
\eea
2-vectors ${\cal{V}}^a$ :
\bea
\label{commv-und}
{\hat{\dot {\cal{V}}}}_{\bar a}-{\dot{\hat {\cal{V}}}}_{\bar
  a} &=&-\A\,\dot{\cal{V}}_{\bar a}+\bra{\frac13\Theta+\Sigma}\hat{\cal{V}}_{\bar a}
+\bra{\Sigma_b+\veps_{bc}\,\Omega^c-\alpha_b}\delta^b{\cal{V}}_a \nn\\
&&+ \A_a \bra{\Sigma_b+\veps_{bc}\,\Omega^c}{\cal{V}}^b + \H\,
\veps_{ab}\,{\cal{V}}^b\ ,
\eea
\bea
\label{commv-deldot}
\delta_a\dot{\cal{V}}_b-\bra{\delta_a{\cal{V}}_b}^{~\cdot}_{\perp}
&=&-\A_{a}\,\dot{\cal{V}}_b+
\bra{\alpha_{a}+\Sigma_a-\veps_{ac}\,\Omega^c}\hat{\cal{V}}_{\bar b}
+\bra{\frac13\Theta-\frac12\Sigma}\bra{\delta_a{\cal{V}}_b+{\cal{V}}_a\,\A_b}
 \nn \\ && +
\bra{\Sigma_{ac}+\Omega\,\veps_{ac}}\bra{\delta^c{\cal{V}}_b+{\cal{V}}^c\,\A_b}
+\frac12\bra{{\cal{V}}_a\,Q_b-N_{ab}\,{\cal{V}}^c\,Q_c}\nn \\ &&
 -\bra{\frac12\phi\,
  N_{ac}+\xi\,\veps_{ac}+\zeta_{ac}}{\cal{V}}^c\,\alpha_b 
  + \H_a\,\veps_{bc}\,{\cal{V}}^c\ , 
\eea
\bea
\label{commv-delhat}
\delta_a\hat{\cal{V}}_b-(\widehat{\delta_a{\cal{V}}_b})^{~\hat{}}_{\perp} &=&
-2\,\veps_{ac}\,\Omega^c\,\dot{\cal{V}}_{\bar b} +
\hatn_a\,\hat{\cal{V}}_{\bar b}
 +\frac12\phi\bra{\delta_a{\cal{V}}_b-{\cal{V}}_a\,a_b}+\bra{\zeta_{ac}
 +\xi\,\veps_{ac}}\bra{\delta^c{\cal{V}}_b-{\cal{V}}^c\,a_b} \nn
 \\ && - 2\bra{\Omega\,\veps_{a[b}+\Sigma_{a[b}}\bra{\Sigma_{c]}
+\veps_{c]d}\Omega^d}{\cal{V}}^c \nn \\&&
-\bras{\bra{\frac12\Sigma-\frac13\Theta}\bra{\Sigma_b
+\veps_{bc}\,\Omega^c}
+\frac12\Pi_b+\E_b}{\cal{V}}_a \nn \\ && + N_{ab}
\bras{\bra{\frac12\Sigma-\frac13\Theta}\bra{\Sigma_c
+\veps_{cd}\,\Omega^d}+\frac12\Pi_c+\E_c}{\cal{V}}^c\ ,
\eea
\bea
\label{commv1d}
\delta_{[a}\delta_{b]}{\cal{V}}^c &=&
\bras{\bra{\frac13\Theta-\frac12\Sigma}^2 -\frac14\phi^2+\frac12\Pi+\E
  -\frac13\mu}{\cal{V}}_{[a}N_{b]}^{~~c} \nn \\
&&
-{\cal{V}}_{[a}\bras{-\bra{\frac13\Theta-\frac12\Sigma}\bra{\Sigma_{b]}^{~~c}
+\Omega\,\veps_{b]}^{~~c}}
+\frac12\phi \bra{\zeta_{b]}^{~~c}+\xi\,\veps_{b]}^{~~c}}
+\frac12\Pi_{b]}^{~~c}+\E_{b]}^{~~c}}
      \nn \\
&& +N_{[a}^{~~c}\bras{-\bra{\frac13\Theta-\frac12\Sigma}
 \bra{\Sigma_{b]d}+\Omega\,\veps_{b]d}}
+\frac12\phi\bra{\zeta_{b]d}+\xi\,\veps_{b]d}}
+\frac12\Pi_{b]d}+\E_{b]d}}{\cal{V}}^d
      \nn \\
&&-\bras{\bra{\Sigma_{[a}^{~~c}
+\Omega\,\veps_{[a}^{~~c}}\bra{\Sigma_{b]d}+\Omega\,\veps_{b]d}}
-\bra{\zeta_{[a}^{~~c}+\xi\,\veps_{[a}^{~~c}}\bra{\zeta_{b]d}
+\xi\,\veps_{b]d}}}{\cal{V}}^d 
	+\veps_{ab}\bra{\Omega\,\dot{\cal{V}}^{\bar c}-\xi\,\hat{\cal{V}}^{\bar c}} ~,
\eea
where we have used both the bar `$~\bar{}$ ' over the index and `$\perp$' to denote 
projection onto the sheet.

Analogous relations for second-rank tensors hold but are more complicated.
\subsubsection{1+1+2 covariant equations}
The key variables of the 1+1+2 formalism of FOG are the irreducible set of geometric variables, 
\be
\{ R, \, \Theta, \,\A, \,\Omega,\,\Sigma,\, \E, \, \H, \, \phi, \, \xi, \, \A_{a}, \, \Omega_{a}, \, \Sigma_{a},\, 
 \alpha_{a}, \,\hatn_{a}, \,  \E_{a}, \,\H_{a},\, \Sigma_{ab},\,\zeta_{ab}, \,\E_{ab},\, \H_{ab} \} ~, 
 \label{splitvar}
\ee
together with the set of irreducible thermodynamic matter variables,
\be
\{\mu^{M},\, p^{M},\, Q^{M},\,\Pi^{M},\,Q^{M}_a,\,\Pi^{M}_a,\,  \Pi^{M}_{ab} \}~, \label{thersplit}
\ee
for a given equation of state. The full 1+1+2 equations for the above 
covariant variables can be obtained by applying the 1+1+2 decomposition procedure to the 
1+3 equations, and in addition, by covariantly splitting the Ricci 
identities for $n^{a}$:
\be
\label{riccisplit}
R_{abc}\equiv 2\na_{[a}\na_{b]} n_c -R_{abcd }n^d=0~,
\ee
where $R_{abcd}$ is the Riemann curvature tensor. By splitting this third-rank tensor using the 
two vector fields $u^{a}$ and $n^{a}$, we obtain the {evolution} equations (along $u^{a}$) 
and {propagation} equations (along $n^{a}$) for $\alpha_a$, $\hatn_a$, $\phi$, $\xi$ 
and $\zeta_{ab}$. The full set of 1+1+2 equations for arbitrary spacetimes is given in 
\cite{Clarkson:2007} .

\section{Perturbations around a Schwarzschild black hole in $f(R)$ gravity}
\label{chp:BH}
In this section we present the complete set of 1+1+2 covariant and gauge invariant evolution, 
propagation and constraint equations linearised around the Schwarzschild background 
in $f(R)$ gravity.
\subsection{Gauge invariance}
In the standard approach to investigating perturbations, any quantity $\cal T $ in the physical manifold $\MM$ 
can be split into a background part ${\cal{T}}_{0}$ on the background manifold $\bar \MM$ and 
a small perturbation $\delta  \cal T$.
\be
\mathcal T  = \mathcal{T}_{0} + \delta  \mathcal T \label{tensorsplit}
\ee
To define the perturbations a gauge choice has to be made. This essentially corresponds to a choice of 
the mapping $\Phi$ between the real spacetime defined by the manifold $\MM$ and the fiducial 
(background) manifold $\bar \MM$. The existence of arbitrary numbers of mappings corresponds to 
the gauge freedom of the theory and herein lies the problem of choosing the best way of constructing this  
mapping or correspondence -  also known as the ``fitting problem" in cosmology \cite{Ellis:1989}. 
If a quantity is invariant under this choice of mapping, then it is gauge invariant.

An alternative definition of gauge invariance is described by the Stewart \& Walker lemma \cite{Stewart:1974}: This states 
that a variable is gauge invariant in $ \MM$ if and only if it either
\begin{enumerate}[i.]
\item {vanishes in $ \bar \MM$ ,}
\item{is a constant scalar in $\bar \MM$,}
\item{is a constant linear combination of products of Kronecker deltas with constant coefficients.} 
\end{enumerate}
The definition of gauge invariance we use here is from the first two options. In this case the mapped 
quantity will be constant regardless of choice of mapping $\Phi$. 

The covariant approach presented here is based on the introduction of a partial frame in 
the tangent space of each point. Once the frame has been chosen, a complete set of 
covariantly defined (i.e., gauge invariant) exact variables, all of which vanish in the background, 
are obtained. These variables make up the equations describing the true spacetime. Since the true 
spacetime lacks the symmetry of the background, there are a number of 
natural choices for the choice of frame vectors and it then follows that one is free to choose 
the frame to work in. Hereafter, the term `frame invariant' refers 
to invariance under the choice of frame vectors. 

\subsection{Schwarzschild background}
The background spacetime we consider is spherically symmetric. Spherically symmetric spacetimes 
are rotationally symmetric about a preferred spatial direction with zero vorticity \cite{vanElst:1996b}. 
Since continuous symmetry of isotropy at each point applies, of all 1+1+2 vectors and tensors vanish 
and the spacetime is described by the covariantly defined scalars:
\be
\{ R, \, \Theta, \,\A, \,\Omega,\,\Sigma,\, \E, \, \H, \,
 \phi,\, \xi, \,\mu^{M},\, p^{M},\, Q^{M},\,\Pi^{M} \} ~.
\ee
 The further constraint that the vorticity 
vanishes $\Omega=\xi = 0$ results in a zero magnetic Weyl curvature scalar $\H = 0$. Thus the variables
\be
\{ R, \, \Theta, \,\A, \,\Sigma,\, \E, \, \phi, \,\mu^{M},\, p^{M},\, Q^{M},\,\Pi^{M} \} ~, 
\ee
fully describe the spherically symmetric spacetime.

If we consider the geometry of a vacuum ($\mu^M= p^M=Q^M=\Pi^M=0$) spherically 
symmetric spacetime, then the set of scalars that describe spacetime reduces to 
\be
\qquad \brac{R, \A, \,\Theta,\,\phi,
 \,\Sigma,\,\E }\ .
\ee 
The condition of staticity implies that $\Theta$ and $\Sigma$ vanish \cite{Nzioki:2014}.\\
If we impose further the conditions 
\bea
&|\fp(0)|<+\infty\,, \qquad |\fpp(0)|<+\infty\,,\qquad|\fppp(0)|<+\infty\,&.\\
&f(0)=0, \qquad R=0, \qquad \fp(0)\ne 0~,&
\eea
the system of equations for the variables reduces to
\bea
\label{schwphi} 
\hat\phi &=& - \,\frac12 \phi^{2} - \E \, ,\\
\label{schwE} 
\hat \E &=& -\, \frac32  \phi \E \, ,\\
\label{schwA}
\hat{\A} &=& -\,\A\bra{ \phi + \A}  \, ,
\eea
together with the constraint:
\bea
\label{schwconstraint}
\E + \A \phi = 0 ~. 
\eea
The parametric solutions for these variables are
\bea
\label{Schw}
\phi=\frac{2}{r}\sqrt{1-\frac{2m}{r}}\;, \qquad \A=\frac{m}{r^2}
\bras{1-\frac{2m}{r}}^{-\frac12}~,
\qquad\E=\frac{2m}{r^3}~,
\eea
where $m$ is the Schwarzschild mass.

\subsection{Linearised field equations}
We now linearise the field equations (evolution, propagation and constraint) as given in \cite{Clarkson:2007} for FOG
\footnote{As a reminder, the thermodynamic quantities in \cite{Clarkson:2007} are derived from the total effective EMT that 
comprises both the standard matter and curvature fluid terms.}
around a Schwarzschild background. The background is 
characterised by the variables $\{ \A,\, \E,\,\phi \}$ and  $\{\hat{\A}, \,\hat{\E},\, \hat{\phi} \}$ which are 
of zeroth-order.  The remaining set of 1+1+2 variables
\bea
\label{firstorder}
\{R,\,\Theta, \,\Sigma, \,\Omega, \,\H, \,\xi, \,\A^{a},\, \Omega^{a}, \,\Sigma^{a}, 
\,\alpha^a,\, \hatn^a,\, \E^{a}, \,\H^{a},\,\Sigma_{ab }, \,\E_{ab},\,\H_{ab},\,\zeta_{ab}\}\,,
\eea
are first-order variables which vanish in the background. These quantities are all of 
${\mathcal O}(\epsilon)$ with respect to the Schwarzschild radius which sets up 
the scale for perturbations for a vacuum spherically 
symmetric spacetime with vanishing Ricci scalar \cite{Nzioki:2014}.
Keeping in mind that gauge invariance holds for the variables \reff{firstorder}, we linearise 
the equations by neglecting the products of these variables along with their derivatives and the dot 
- $`\dot{\phantom{x}}$' and 
delta - $`\delta$' derivatives of $\brac{\A,\, \E,\,\phi}$ to obtain:

{\it Evolution equations}: 
 \bal
 \label{dotphil2}
\dot{\phi} &= \bra{\frac23 \Theta - \Sigma}\bra{\A-\frac12 \phi}
+\delta_a\alpha^a +\frac{\fpp_{0}}{\fp_{0}}\, (\A \,\dot{R}-\dot{X}) ~,
\eal

\bal
\label{dotxinl2}
\dot{\xi} &= \bra{\A-\frac12\phi}\Omega
+\frac12\,\veps_{ab}\delta^a\alpha^b +\frac12 \H ~,
\eal

 \bal
\dot\Omega&=\frac12\veps_{ab}\delta^a\A^b+\A\,\xi~,
\eal

\bal
\label{dotSigSnl2}
\dot\Sigma - \frac23 \dot \Theta &= -\phi\, \A - \delta_a\A^a - \E
- \frac{\fpp_{0}}{2\fp_{0}}\bra{\delta^2R+\bra{  \phi+2\A } X
-2\ddot{R}}~,
\eal

\bal
\label{dotEl2}
\dot\E &= \bra{\frac32 \Sigma - \Theta} \E+ \veps_{ab} \delta^a\H^b
+\phi\, \A\,\frac{\fpp_{0}}{2\fp_{0}}\,\dot{ R} ~, 
\eal

\bal
\dot\H&=-\,\veps_{ab}\delta^a\E^b-3\xi\,\E~,
\eal

\bal
\dot\Sigma_{\bar a}-\veps_{ab} \dot{\Omega}^b 
&=  \delta_a\A+\bra{\A-\frac12\phi}\A_a   -\E_a 
 +\frac{\fpp_{0}}{2\fp_{0}}\bra{\delta_{a} X
 - \frac12 \phi \, \delta_{a} R}~,
\eal

\bal
\dot\E_{\bar a} + \frac12 \veps_{ab} \hat\H^b  =&\,
	 \frac34\, \E  \bra{\veps_{ab}\Omega^b + \Sigma_a-2 \alpha_a}
	 - \bra{\frac14 \phi+ \A} \veps_{ab}\H^b  \nn\\
	& +\, \frac34\,\veps_{ab} \delta^b\H
	+\frac12\,\veps_{bc} \delta^b\H^c{}_a ~,
\eal 

\bal
\dot\H_{\bar a} &=
   -\,\frac32 \E\, \veps_{ab} \A^b -\frac12\veps_{ab} \delta^b\E 
 -\frac12 \bra{\phi - 2\A} \veps_{ab} \E^b
   + \veps_{c\lb d} \delta^d \E_{a\rb}^{~~c} 
 - \E \,\frac{\fpp_{0}}{4\fp_{0}} \,\veps_{ab}\delta^b R~,
\eal

 \bal
  \label{dotzetanl2}
\dot\zeta_{\lb ab\rb}&=\bra{\A-\frac12\phi}\Sigma_{ab}
+\delta_{\lb a}\alpha_{b\rb} 
 -\veps_{c\lb a}\H_{b\rb}^{~~c}~,
\eal

 \bal
\dot\Sigma_{\lb ab\rb}&=\delta_{\lb a}\A_{b\rb} 
    +\A\,\zeta_{ab}-\E_{ab}
    +\frac{\fpp_{0}}{2 \fp_{0}}\, \delta_{\lb a} \delta_{b \rb} R~,
\eal

\bal
\label{divSigmanl2}
\frac{\fpp_{0}}{\fp_{0}}\, \delta_{a} \dot{R} &=\delta_a\Sigma
	-\frac23\delta_a\theta
	+2\,\veps_{ab}\delta^b\Omega +2\,\delta^b\Sigma_{ab}
	+\phi\bra{\Sigma_a+\veps_{ab}\Omega^b}
	+2\veps_{ab}\H^b   ~.
\eal
{ \it Propagation equations}:
\bal
\label{hatphil2}
\hat\phi&=   -\,\frac12\phi^2 - \E + \delta_a a^a 
- \frac{\fpp_{0}}{2\fp_{0}} \bra{2\hat{X}+\phi\,X+\delta^2R }~,
\eal

\bal
\label{hatxinl2}
\hat\xi&=-\,\phi\,\xi+\frac12\veps_{ab}\delta^a \hatn^b~,
\eal

\bal
\label{hatOmSnl2}
\hat\Omega&= -\,\delta_a\Omega^a+\bra{\A-\phi}\Omega ~,
\eal

\bal
\label{hatAl2}
\hat\A - \dot \Theta &=- \delta_a \A^a  -\bra{\A+\phi} \A
+ \frac{\fpp_{0}}{2\fp_{0}}\bras{3\ddot{R}-\delta^2R 
- \hat{X} - (3\A+\phi) X} ~,
\eal

\bal
\hat\Sigma - \frac23 \hat \Theta &= -\, \frac32 \phi \,\Sigma
 - \delta_a\Sigma^a - \veps_{ab}\delta^a \Omega^b
 +  \frac{\fpp_{0}}{\fp_{0}}   \bra{ \dot{X}-\A \,\dot{R}  }~,
\eal

\bal
\label{hatEl2}
\hat \E&=-\frac32 \phi \,\E  -\,\delta_a\E^a  
- \E\, \frac{\fpp_{0}}{2\fp_{0}} \,  X ~,
\eal

\bal
\hat\H&= -\,\delta_a\H^a -\frac32\phi\,\H -3\E\, \Omega~,
\eal

\bal
\label{hatalphanl2}
\dot a_{\bar a}-\hat\alpha_{\bar a}&=
    \bra{\frac12\phi+\A}\alpha_a
    -\bra{\frac12\phi-\A}\bra{\Sigma_a+\veps_{ab}\Omega^b}
  +\veps_{ab}\H^b +\frac{\fpp_{0}}{2\fp_{0}}\,\delta_{a} \dot{R}~,
\eal

\bal
\hat\Sigma_{\bar a}-\veps_{ab}\hat\Omega^b=&\,\frac12\delta_a\Sigma
    +\frac23\delta_a\theta - \veps_{ab}\delta^b\Omega
    -\frac32\phi\,\Sigma_a \nn \\
    &+\,\bra{\frac12\phi+2\A}\veps_{ab}\Omega^b
    -\delta^b\Sigma_{ab} +  \frac{\fpp_{0}}{\fp_{0}} \,\delta_{a} \dot{R}~,
\eal

 \bal
 \label{hatAanl2}
\hat{\A}_a- 2\dot{\Sigma}_a &=
	-\,\delta_a \A-2\bra{\A-\frac14 \phi}\A_a 
	-\A \, \hatn_a + 2 \E_a 
	- \frac{\fpp_{0}}{\fp_{0}}\, \bra{\delta_{a} X
 - \frac12 \phi \, \delta_{a} R}~.
\eal

\bal
\hat\E_{\bar a} &= \frac12\delta_a\E
    -\delta^b\E_{ab} -\frac32 \E\, a_a
    -\frac32\phi \, \E_a 
    +\E  \, \frac{\fpp_{0}}{4\fp_{0}} \,  \delta_a R~,
\eal

 \bal
 \label{hatHnl2}
\hat\H_{\bar a} &=
	\frac12\delta_a\H-\delta^b\H_{ab}
  	  +\frac32 \E \bra{\Omega_a-\veps_{ab}\Sigma^b }
  	  -\frac32\phi\,\H_a  ~,
\eal

\bal
\label{hatzetanl2}
\hat\zeta_{\lb ab\rb}&=-\,\phi\,\zeta_{ab}
	+\delta_{\lb a} a_{b\rb } -{\cal E}_{ab} 
	- \frac{\fpp_{0}}{2\fp_{0}} \,  \delta_{\lb a} \delta_{b \rb} R~,
\eal

\bal
\hat\Sigma_{\lb ab\rb}&=\delta_{\lb a}\Sigma_{b\rb} 
-\veps_{c\lb a}\delta^c\Omega_{b\rb}
    -\frac12\phi\,\Sigma_{ab} -\veps_{c\lb a}\H_{b\rb}^{~~c}~,
\eal

\bal
\dot\E_{\lb ab\rb}-\veps_{c\lb a}\hat\H_{b\rb}^{~~c}&=
	-\,\veps_{c\lb a}\delta^c\H_{b\rb}
	+\bra{\frac12\phi+2\A}\veps_{c\lb a}\H_{b\rb}^{~~c}
	 -\frac32 \E\, \Sigma_{ab}~,
\eal

\bal
\dot\H_{\lb ab\rb}+\veps_{c\lb a}\hat\E_{b\rb}^{~~c} &=
	\veps_{c\lb a}\delta^c\E_{b\rb}
	+\frac32 \E\,\veps_{c\lb a}\zeta_{b\rb}^{~~c}
	-\bra{\frac12\phi+2\A}\veps_{c\lb a}\E_{b\rb}^{~~c}~.
\eal

\bal
\label{divzetanl2}
  \frac{\fpp_{0}}{2\fp_{0}}\bra{  \delta_{a} X
    - \frac12 \phi \delta_{a} R } &= -\frac12\delta_a\phi
    +\veps_{ab}\delta^b\xi+\delta^b\zeta_{ab}
   -\E_a~,
\eal
\\ \\
{\it The trace equation}:
\bal
\label{Curvetracenl2}
 \fpp_{0} ( \hat{X}  - \ddot{R}) = \frac13 R \, \fp_0 
 -\fpp_{0} \bras{\delta^{2} R +  \bra{\phi +\A} X} ~.
\eal
\\ \\
{\it Constraint equations}: 
\bal
\delta_a\Omega^a+\veps_{ab}\delta^a\Sigma^b&=
	\bra{2\A-\phi}\Omega+\H~,
	\label{divOmeganl2}
\eal \\
In the above equations, $X=\hat{R},~ \fp_0 = \fp(0)$ and  $ \fpp_0 = \fpp(0)$.
\subsection{Gauge invariant variables}
Not all the set of covariant equations in the previous section are gauge invariant 
due to the isolated zeroth-order background terms that appear in them. To fix this, we define three 
key variables by taking the angular derivatives of the background variables $\{\E,\, \phi, \,\A\}$, 
\bea
W_a &=& \delta_a \E ~,
\\
Y_a &=& \delta_a \phi~, 
\\
Z_a &=& \delta_a \A ~.
\eea
These new variables vanish in background and are therefore gauge invariant. Applying the 
commutation relations \reff{comm2} and \reff{comm3} and substituting for the 
subsequent equations, we obtain the following linearised 
propagation and evolution equations for these new variables:
\bea
\label{newEdot} 
\dot W_a &=&\frac32 \phi \,\E \bra{\alpha_a + \Sigma_a - \veps_{ab} \Omega^b}
+ \frac32 \E\bra{ \delta_a \Sigma - \frac23 \delta_a \Theta} + \veps_{bc} \delta_a \delta^b \H^c 
+ \A\, \phi \,\frac{\fpp_{0}}{2\fp_{0}}\, \delta_{a} \dot{R}~, \\
\label{newphidot}
\dot Y_a &=& \bra{\frac12 \phi^2 + \E}\bra{\alpha_a + \Sigma_a - \veps_{ab} \Omega^b}
+ \delta_a\delta_c \alpha^c 
+\bra{\frac12 \phi - \A}\bra{\delta_a \Sigma - \frac23 \delta_a \Theta}  
+\frac{\fpp_{0}}{\fp_{0}}\,  \bra{\A \, \delta_{a} \dot{R}  - \delta_{a} \dot{X} }~,   \\
\label{newEhat}
 \hat W_a &=& -\,2 \phi \,W_a - \frac32 \E\, Y_a 
+ \frac32 \phi \,\E \,\hatn_a - \delta_a \delta_b \E^b
-  \E\, \frac{\fpp_{0}}{2\fp_{0}}\, \delta_a X~,  \\
\label{newphihat}
\hat Y_a &=& -\, W_a - \frac32 \phi\, Y_a +\bra{\frac12 \phi^2 + \E} a_a  
+ \delta_a \delta_b a^b -  \frac13 \delta_a R  \nn \\
&&+\,\frac{\fpp_{0}}{\fp_{0}} \bras{ \bra{\A +\frac12 \phi} \delta_a X 
+\frac12 \bra{\E - \frac14 \phi^2} \delta_a R 
+ \frac12 \delta^2\delta_a R- \delta_a \ddot{R} } ~,  \\
\label{newAhat}
\hat Z_a &=& -\bra{\frac32 \phi + 2 \A} Z_a - \A\, Y_a + \A \bra{\phi + \A} a_a
+\delta_a \dot \Theta 
- \delta_a \delta_b \A^b +\frac{\fpp_{0}}{\fp_{0}}\bra{ \delta_{a} \ddot{R} 
- \A\, \delta_{a} \dot{X} } ~.  
\eea
These equations add no new information to what has already been given in 
the previous section however, since they are gauge invariant,
we can replace the equations \reff{dotEl2}, \reff{dotphil2}, \reff{hatEl2},  
\reff{hatphil2} and \reff{hatAl2} with \reff{newEdot}, \reff{newphidot}, \reff{newEhat}, 
\reff{newphihat} and \reff{newAhat} respectively.\\
The following additional constraints are obtained by applying the commutation 
relation \reff{comm3} to the new variables  \reff{newEdot}-\reff{newAhat} ,
\bea
\label{Wconst}
\veps_{ab} \delta^a W^b &=& 3 \phi\, \E \,\xi ~,
\\
\label{Yconst}
\veps_{ab} \delta^aY^b&=& \bra{\phi^2 + 2 \E} \xi ~,
\\
\label{Zconst}
\veps_{ab} \delta^aZ^b &=& 2\A \bra{\phi + \A} \xi ~.
\eea
It is also useful to replace \reff{dotSigSnl2} with
 \bea
 \label{WYZ}
\delta_a\dot\Sigma-\frac23\delta_a\dot\theta&=&-\,W_a-\A \,Y_a
-\phi\, Z_a-\delta_a\delta_b\A^b 
- \frac{\fpp_{0}}{2\fp_{0}}\bras{\delta^2\delta_aR 
-2\delta_a\ddot{R} \phantom{\frac11} 
 +\bra{\E - \frac14 \phi^2}\delta_aR+\bra{  \phi+2\A } \delta_aX}~.
\eea

\subsection{Commutation relations}
The following are the relevant commutation relations for the derivatives of 
first-order scalar, vector and tensor quantities, ${ \cal{T}}$: 

{\it Scalars}:
\bea
{\dot{\hat{ \cal{T}}}}-\hat{\dot { \cal{T}}}&=&\A\,\dot{ \cal{T}}~,
\\
\delta_a\dot{ \cal{T}}-\bra{\delta_a{ \cal{T}}}^\cdot &=&0~,
\\
{\delta_a\hat{ \cal{T}}}-{\widehat{\bra{\delta_a{ \cal{T}}}}} &=&\frac12\phi\,\delta_a{ \cal{T}}~,
\\
\delta_{[a}\delta_{b]}{ \cal{T}}&=&0~;
\eea
{\it Vectors}:
\bea
{\dot{\hat { \cal{T}}}}_{\bar{a}}-{\hat{\dot { \cal{T}}}}_{\bar{a}}&=&\A\,\dot{ \cal{T}}_{\bar{a}}~,
\\
\delta_{[a}\delta_{b]}{ \cal{T}}_c&=&\bra{\frac14\phi^2-\E}N_{c[a}{ \cal{T}}_{b]}~;
\label{vectcomm}
\eea
{\it Tensors}:
\bea
{\dot{\hat{ { \cal{T}}}}}_{\brac{ab}}-{\hat{\dot {{ \cal{T}}}}}_{\brac{ab}}&=&\A\,\dot{ \cal{T}}_{\brac{ab}}~,
\\
\delta_{[a}\delta_{b]}{ \cal{T}}_{cd}&=&\bra{\frac14\phi^2
-\E}\bra{N_{c[a}{ \cal{T}}_{b]d}+N_{d[a}{ \cal{T}}_{b]c}}~.
\eea
\subsection{Harmonic decomposition}
In order to solve the equations, it is standard procedure to decompose the first order variables 
harmonically (see, \cite{Hawking:1966, Bardeen:1980}). The perturbations can be described 
by a linear system of ODEs by introducing spherical and time harmonics. 
\subsection{Spherical harmonics}
 We perform a decomposition of first order perturbations into scalar, vector and tensor modes in 
 analogy with the FLRW models \cite{Bruni:1992, Ellis:1998}. The perturbations of the Schwarzschild 
 geometry fall into two distinct classes based on how they transform on the surfaces of 
 spherically symmetry: even (electric) and odd (magnetic) modes 
 \footnote{Alternatively, as first presented in Chandrasekhar's book \cite{Chandrasekhar}, 
 odd perturbations are called {axial} and even perturbations are called polar.}. 
 Given the spherical symmetry of the background, we can naturally use spherical harmonics 
 to expand the first order quantities. This being the case, the scalars can be expanded as a sum 
 of even modes and the vectors and tensors can be expanded in sums over both the even and 
 odd modes. Moreover, the angular derivatives appearing in the equations are effectively replaced by a 
 harmonic component of the derivative. 
 The presentation in this section follows \cite{Clarkson:2003} where the harmonics were introduced 
 in a covariant manner.
 
We introduce the set of dimensionless spherical harmonic functions $Q=Q^{(\ell,m)}$, 
with $m=-\ell,\cdots,\ell$, defined on the background as eigenfunctions 
of the spherical Laplacian operator such that 
\be
\label{SH}
\delta^2 Q =-\, \frac{\ell(\ell+1)}{r^2} \,Q~.
\ee
The function $Q$ is defined in order to be covariantly constant along 
$u^a$ and $n^a$, 
\be
\hat{Q} = 0 = \dot{Q}~.
\ee
The function $r$ is, up to an arbitrary constant, covariantly defined by
\be
\frac{\hat r}{r}\, = \frac12\,\phi\ , \qquad {\dot r}= 0 =\delta_a\, r \ , 
\label{rdefder}
\ee
and gives a natural length scale to the spacetime as seen when $r$ is defined as
\be
r\equiv\bra{\frac14\phi^2-\E}^{-1/2}~.
\label{rdef}
\ee
We stress that these relations and harmonics are used in 
expanding gauge invariant first-order quantities only.

We now look successfully at the expansion of first order scalars, vectors and 
tensors in spherical harmonics and the replacements which must be made in 
the equations. 

\subsubsection*{Scalar harmonics}
We can now define the harmonic expansion of any first order
scalar ${\Psi}$ in terms of the functions $Q$ as
\be
{\Psi}=\sum_{\ell=0}^{\infty}\sum_{m=-\ell}^{m=\ell} {\Psi}\S^{(\ell,m)}
Q^{(\ell,m)} = {\Psi}\S \,Q,
\ee
where from now on we drop the sum over $\ell$ and $m$ (implicit in the last equality) 
in the harmonic expansions hereafter. 
We use the
subscript $\mathsf{S}$ to indicate that a scalar spherical harmonic expansion 
has been made. 	

The replacements  which must be made for scalars when expanding the equations 
in spherical harmonics are
\bea
 {\Psi} &=& {\Psi}\S\, Q ~,                
  \\
 \delta_a {\Psi} &=& r^{-1}{\Psi}\S \,Q_a~,
 \\
 \veps_{ab}\delta^b {\Psi} &=& r^{-1}{\Psi}\S\, \bar Q_a ~.
\eea

\subsubsection*{Vector harmonics}
The vector harmonics can be either of {even} (electric) or {odd} 
(magnetic) parity. The {even} parity vector spherical harmonics for 
$\ell\geq1$ we will define as
\be
Q_a^{(\ell)}=r\,\delta_a Q^{(\ell)} 
\label{evenvector}
\ee
where $Q_a$ is covariantly constant along $u^a$ and $n^a$ 
\be
\hat Q_a=0=\dot Q_a~.
\ee

The vector harmonic \reff{evenvector} is defined as an eigenfunction of the 
spherical Laplacian operator:	 
\be
\delta^2 Q_a=\bra{1-\ell\bra{\ell+1}}r^{-2} Q_a~,
\ee
and satisfies the properties 
\bea
\delta^aQ_a &=& -\ell\bra{\ell+1}r^{-1} Q~,\\
\veps_{ab}\delta^a  Q^b &=& 0~.
\eea
Similarly, we define {odd} parity vector spherical harmonics as
\be
\bar Q_a^{(\ell)}=r\,\veps_{ab}\delta^b Q^{(\ell)}~~~\Rightarrow
~~~\hat{\bar{Q}}_a=0=\dot{\bar{Q}}_a~,~~~\delta^2\bar
Q_a=\bra{1-\ell\bra{\ell+1}}r^{-2}\bar Q_a~,
\label{oddvector}
\ee
$\bar Q_a$ being a solenoidal vector,
\be
\delta^a\bar Q_a=0~,
\ee
and satisfies the property
\be
\veps_{ab}\delta^a\bar Q^b=\ell\bra{\ell+1}r^{-1} Q~.
\ee
Note that $Q_a$ and $\bar Q_a$ are parity inversions of one another other 
\be
\bar Q_a=\veps_{ab}Q^b\Leftrightarrow Q_a=-\,\veps_{ab}\bar Q^b~,
\ee
where $\veps_{ab}$ is a parity operator. \\
Since the even and odd vector harmonics are orthogonal: $Q^a\,\bar Q_a=0$ (for each $\ell$), 
then any first-order vector ${\Psi}_a$ may be expanded in terms of 
these harmonics as
\be
{\Psi}_a=\sum_{\ell=1}^{\infty} {\Psi}^{(\ell)}\V \,Q_a^{(\ell)}+\bar
{\Psi}^{(\ell)}\V\,\bar Q_a^{(\ell)}={\Psi}\V\, Q_a+\bar {\Psi}\V\,\bar Q_a~.
\ee
where the $\mathsf{V}$ indicates that a vector spherical harmonic expansion 
has been made.\\
As in the scalar case, the replacements to be made for vectors when 
expanding the equations in spherical harmonics are
\bea
 {\Psi}_a& =&{\Psi}\V \,Q_a+\bar {\Psi}\V\,\bar Q_a ~,                
  \\
 \veps_{ab} {\Psi}^b&=& -\, \bar {\Psi}\V\, Q_a+{\Psi}\V\,\bar Q_a    ~,
 \\
\delta^a{\Psi}_a&=&-\,\ell\bra{\ell+1}r^{-1}{\Psi}\V \,Q  ~,
\\
\veps_{ab}\delta^a{\Psi}^b&=&\ell\bra{\ell+1}r^{-1}\bar {\Psi}\V\, Q ~,
\\
 \delta_{\lb a}{\Psi}_{b\rb}&=&r^{-1}\bra{{\Psi}\V\, Q_{ab}
 -\bar {\Psi}\V \,\bar Q_{ab}} ~,
\\
\veps_{c\lb a}\delta^c {\Psi}_{b\rb}&=& r^{-1}\bra{\bar {\Psi}\V\, Q_{ab}
+ {\Psi}\V \,\bar Q_{ab}}~.
\eea
\subsubsection*{Tensor harmonics}
We define even and odd parity tensor spherical harmonics for $\ell\geq2$ as
\bal
Q_{ab} = r^2\,\delta_{\lb a}\delta_{b\rb}Q,~~\Rightarrow& ~~\hat Q_{ab}=0=\dot
Q_{ab},~~~\delta^2Q_{ab}=\bras{\phi^2-4\E-\ell\bra{\ell+1}r^{-2}}Q_{ab}~,\\
\bar Q_{ab}= r^2\,\veps_{c\lb a}\delta^c\delta_{b\rb}Q~,~~\Rightarrow& ~~\hat{\bar
Q}_{ab}=0=\dot{\bar Q}_{ab},~~~\delta^2\bar
Q_{ab}=\bras{\phi^2-4\E-\ell\bra{\ell+1}r^{-2}}\bar Q_{ab},
\eal
and posses the same orthogonal and parity property
\ber
Q_{ab}\,\bar Q^{ab}&=& 0~,\\
Q_{ab}=-\veps_{c\lb a}\bar Q_{b\rb}^{~~c}&\Leftrightarrow&
\bar Q_{ab}=\veps_{c\lb a} Q_{b\rb}^{~~c}~,
\ear
as the vector case. Any first-order tensor ${\Psi}_{ab}$ can be 
expanded in terms of these harmonics as
\be
{\Psi}_{ab}=\sum_{\ell=2}^{\infty} {\Psi}\T^{(\ell)}\,Q_{ab}^{(\ell)}+\bar
{\Psi}\T^{(\ell)}\,\bar Q_{ab}^{(\ell)}={\Psi}\T\, Q_{ab}+\bar {\Psi}\T\,\bar Q_{ab}~.
\ee
For the tensors, the following replacements must be made when expanding the 
equations in spherical harmonics:
\bea
 {\Psi}_{ab}&=&{\Psi}\T \,Q_{ab}+\bar {\Psi}\T\,\bar Q_{ab} ~,                
  \\
 \veps_{c\lb a}{\Psi}_{b\rb}{}^{c}&=& -\,\bar {\Psi}\T\, Q_{ab}
 +{\Psi}\T\,\bar Q_{ab}    ~,
 \\
\delta^b {\Psi}_{ab}&=&\bras{1- \frac12\ell (\ell+1)} r^{-1}\bra{{\Psi}\T\, Q_{a}
-\bar {\Psi}\T\,\bar Q_{a}} ~,
\\
\veps_{c\lb d}\delta^d{\Psi}_{a\rb}{}^{c}&=&- \,\bras{1- \frac12\ell (\ell+1)} r^{-1}
\bra{\bar {\Psi}\T\, Q_{a}+ {\Psi}\T\,\bar Q_{a}} ~.
\eea
\subsubsection*{Odd and even parity perturbations}
Expanding the perturbations into spherical harmonics, leads to two independent 
set of equations with the following variables: 

\emph{Odd} perturbations~:
\bal
\vo\equiv& \{\bar\E\T,~\H\T,~\bar\Sigma\T,~\bar\zeta\T\}~,\nn\\
 ~&\{\bar\E\V,~\H\V,~\bar\Sigma\V,~\Omega\V,~\bar\A\V,~\bar\alpha\V,
  ~\bar\hatn\V,
 ~\bar X\V,~\bar Y\V,~\bar Z\V\}~,\nn\\
 ~&\{\H\S,~\Omega\S,~\xi\S\}~;
\label{oddvars}
\eal
\emph{Even} perturbations~:
\bal
\ve\equiv\ &\{\E\T,~\bar\H\T,~\Sigma\T,~\zeta\T\}~,\nn\\
 ~&\{\E\V,~\bar\H\V,~\Sigma\V,~\bar\Omega\V,~\A\V,~\alpha\V, ~\hatn\V,
 ~ X\V,~ Y\V,~ Z\V\}~,\nn\\
 ~&\{\Sigma\S,~\theta\S~R\S\}~;
 \label{evenvars}
\eal
 We see in the equations that `parity switching' occurs 
between some sets of variables where certain terms always 
appear alongside the factor `$\veps_{ab}$' relative to other variables 
(e.g., $\H_{ab}$ and $\Omega^a$ appear alongside `$\veps_{ab}$' relative 
to the variables $\E_{ab}$ and $\Sigma^a$, respectively).

\subsection{Time harmonics}
\label{secTime}
Since the background is static, we can resolve the perturbations into temporal harmonics. 
We do this by performing a Fourier analysis of the time derivatives of the first order quantities 
by decomposing them into their Fourier components. This corresponds  
to assuming a harmonic time dependence $e^{i\omega\tau}$ for the first order variables. 

We define the time harmonic function $T^{(\omega)}$ in the background by
\be
\dot T^{(\omega)}=i\,\omega \,T^{(\omega)}, \quad \hat
T^{(\omega)}=0=\delta_aT^{(\omega)}; \quad \dot\omega=0=\delta_a\omega~.
\ee
From the commutation relation between the $dot$- `$^{.}$' and $hat$- `$~\hat{}~$' derivatives the 
above-defined time harmonic must satisfy
\be
\hat{\dot T}+\A\,\dot T=0~,
\ee
which in turn implies
\be
\hat\omega=-\A\,\omega~,
\label{omegahat}
\ee
in the background. 

Integrating \reff{omegahat} in terms of $r$, gives
\be
\omega= \sigma\bra{1-\frac{2m}{r}}^{-1/2}=\frac{2\sigma}{\phi \,r}~,
\ee
where $\sigma$ is a  constant. Then any first order variable $\Psi$ in the equations may be expanded 
as
\be
\Psi=\sum_\omega \Psi^{(\omega)} T^{(\omega)}=\Psi^{(\omega)} T^{(\omega)}~,
\ee
and the dot - `$^{.}$' derivatives of these first order quantities can be replaced by 
factors of $i \omega$.

\section{The Regge\,-Wheeler equation} 
\label{sec:RW}
In GR, the gravitational perturbations of Schwarzschild black holes are governed by a 
single second-order wave equation, namely the Regge\,-Wheeler equation \cite{Regge:1957},
describing the odd perturbations and the Zerilli equation \cite{Zerilli:1970} describing 
the even perturbations. Both the equations satisfy a Schr\"{o}dinger-like equation and it was 
demonstrated in \cite{Chandrasekhar:1975} that the effective potentials of these equations 
have the same spectra. The aim of this section is to perform an analysis of the perturbation of the
Schwarzschild black hole in $f(R)$ gravity and find a reduced set of master variables 
which obey a closed set of wave equations for these theories. 

\subsection{Gravitational perturbations}
If we consider very large distances from the source ($\A=\phi=0$), the gravitational perturbations 
should be well approximated by a plane wave, with $n^a$ lying in the direction of propagation. 
On imposing the condition that $R$ vanishes at infinity, the plane gravitational waves are 
described by the 1+1+2 transverse-traceless tensors $\E_{ab},~\H_{ab}, ~ \Sigma_{ab}$ and 
$\zeta_{ab}$ only, as in GR. Otherwise there is coupling with the scalar waves which can 
produce other scalar and vector modes. The tensors $\E_{ab}$ and $\H_{ab}$ represent the 
tidal and gravitational waves effects in analogy with the propagation of electromagnetic waves. 
However, the wave equations for these two tensors do not close in the general frame.

If we now consider the general case, apart from the four TT tensors, a number of other 
TT tensors can be constructed from the $\delta$- derivatives of vectors and scalars, 
for example, $\delta_{\{a}W_{b\}}, ~ \delta_{\{a}\hatn_{b\}}, 
~ \delta_{\{a}\delta_{b\}}\Omega$, etc. The wave equations for these tensors can 
be calculated by applying the wave operator 
$\ddot{\Psi}_{\lb ab\rb}-\hat{\hat {\Psi}}_{\lb ab\rb}$ to that tensor $\Psi_{ab}$ 
\cite{Clarkson:2003}. The aim here is to calculate all such possible wave equations 
involving these tensors and systematically eliminating unwanted terms until a closed 
equation is obtained. In particular, calculating the wave operator for $\zeta_{ab}$ and 
$\delta_{\lb a}W_{b\rb}$, we notice that they contain similar terms.

We consider the case of the wave operator for $\zeta_{ab}$, that is, 
$\ddot \zeta_{\lb ab\rb}-\hat{\hat \zeta}_{\lb ab\rb}$, where we apply the following steps:
\begin{itemize}
\renewcommand{\labelitemi}{$-$}
 \item{Take the dot- derivative across \reff{dotzetanl2}, for which the resulting evolution 
equations are substituted.} 
 \item{Substitute for $a_a$ from \reff{newEhat} and $\alpha_a$ from \reff{newEdot} (while 
 utilising the constraints \reff{Wconst}, \reff{divSigmanl2},\reff{Zconst},  \reff{divzetanl2} 
 and \reff{Yconst} to substitute for $\xi,~\Sigma, ~Z_{a}$ even $Y_{a}$ and odd 
 $Y_{a}$ respectively).}
\end{itemize}
What follows is an expression consisting of only $\delta_{\lb a}W_{b\rb}$ and 
$\zeta_{ab}$, for the odd harmonics and $\delta_{\lb a}X_{b\rb},~\zeta_{ab}$ and 
 $\delta_{\lb a}\delta_{b\rb}R$ for the even harmonics. We can recast this result 
 as the wave equation,  
\be
\ddot M_{\lb ab\rb}-\hat{\hat M}_{\lb ab\rb}-\A\,{\hat M}_{\lb ab\rb}
+\bra{\phi^2 + \E}M_{ab}-\delta^2 M_{ab} =0~,
\label{RMtensorwave}
\ee
where we have introduced the dimensionless, gauge-invariant, frame-invariant, 
transverse-traceless tensor $M_{ab}$ defined as
\be
M_{ab}=\frac12\phi \,r^2\,\zeta_{ab}-\frac13r^2\,\E^{-1}\,\delta_{\lb a}W_{b\rb}
+ \frac{\fpp_0}{3\,\fp_0}\,r^2\,\delta_{\lb a}\delta_{b\rb}R~.
\label{Mab}
\ee
The even part of \reff{Mab} is coupled to the curvature term and as a result we 
have to include the trace equation \reff{Curvetracenl2} to achieve 
closure. On the other hand, the curvature term vanishes for the odd part of $M_{ab}$
and this leaves the tensor in exactly the same form as in the GR case 
\cite{Clarkson:2003}.

We can expand \reff{RMtensorwave} into scalar harmonics as
\bea
\ddot { M}-\hat{\hat{  M}}-\A\,\hat { M}
+\bras{\frac{\ell\bra{\ell+1}}{r^{2}}+3\E}{ M}=0~,\label{RW}
\eea
where we let $M=\{M\T,\,\xbar M\T\}$.
In appropriate coordinates the wave equation \reff{RW} is the {\it Regge\,-Wheeler
equation}. Both the odd and even parity parts of $M_{ab}$ satisfy the 
same wave equation \reff{RW}. 

We convert to the parameter $r$ using \reff{hatr}, the time harmonics in \reff{RW} 
and the fact that hat derivative of any scalar $K$ for a static spacetime \cite{Betschart:2004} 
is
\be 
\hat{M}=\frac{1}{2}\,r\,\phi\frac{d M}{d r}\,,
\label{hatr} 
\ee   
to obtain
\be
\kappa^2 M-\frac{2m}{r^2}\,\bras{\frac{2m-r}{r}}  \frac{dM}{dr}
+\bra{ \frac{2m-r}{r}}^2 \frac{d^2M}{dr^2}
+\bra{\frac{2m-r}{r}}\bras{\frac{\ell\bra{\ell+1}}{r^2}-\frac{6m}{r^3}}M=0~.
\label{RWr} 
\ee
We then make a change to the `tortoise' coordinate $r_*$, which is related to $r$ by
\be
r_*=r+2m\,\ln\bra{\frac{r}{2m}-1}~,
\label{tortoise}
\ee
thus, \reff{RWr} can be written in the form
\be
\bra{\frac{d^2}{dr_*^2}+\kappa^2-V\T}M= 0~,
\label{schroedTens}
\ee
with the effective potential $V\T$  
\be
V\T=\bra{1-\frac{2m}{r}}\bras{\frac{\ell\bra{\ell+1}}{r^2}
-\frac{6m}{r^3}}~,
\label{Gravitypot}
\ee 
which is the Regge\,-Wheeler potential for gravitational perturbations. 

\subsection{Scalar perturbations}
\label{secScalar}
The trace equation \reff{Curvetracenl2}, which is a wave equation in the Ricci scalar 
$R$, corresponds to scalar modes that are not present in standard 
GR but occur in $f(R)$ theories of gravity due to the extra scalar 
degree of freedom. The equation constitutes the same generalised 
Regge\,-Wheeler equation for massive scalar perturbations on a LRS 
background spacetimes in GR with 
\be
U^2= \frac{\fp_0}{3\,\fpp_0}~,
\label{effmass}
\ee
as the effective mass of the scalar.

To obtain the familiar Regge\,-Wheeler equation we first rescale $R$ as 
$R=r^{-1}\,\R$ and use \reff{schwphi} and \reff{rdefder} to rewrite equation 
\reff{Curvetracenl2} in the form 
\be
\ddot{\R} - \hat{\hat{\R}} -\A\, \hat{\R}
- \bra{\E - U^2+ \delta^2 } \R= 0~.
\label{Reggescaled}
\ee
Proceeding as in the previous case, we introduce scalar spherical harmonics 
to \reff{Reggescaled} resulting in 
\be
\ddot{\R}\S - \hat{\hat{\R}}\S - \A\, \hat{\R}\S
- \bras{\E -\ti{U}^2- \frac{\ell(\ell+1)}{r^2}} \R\S= 0~.
\label{traceharm}
\ee
where $\ti{U}^2 = \f_1/(3\,\f_2) $ with $\f_1$ and $\f_2$ as constants. \\
Converting to the parameter $r$ and then the tortoise coordinate, we get
\be
\bra{\frac{d^2 }{dr_*^2} 
+ \kappa^2-V\S } \R =0~,
\label{schroedScal}
\ee
where 
\be
V\S=\bra{1-\frac{2m}{r}}\bras{\frac{\ell\bra{\ell+1}}{r^2}
+\frac{2m}{r^3}+\ti{U}^2}~.
\label{Scalpot}
\ee 
The expression \reff{Scalpot} is the Regge\,-Wheeler potential for the scalar perturbations. 
\subsection{Potential profile}
\label{secPotential}
The form of the wave equations \reff{schroedTens} and \reff{schroedScal} 
describing black hole perturbation is similar to a one dimensional 
Schr\"{o}dinger equation and hence their potentials correspond to a single 
potential barrier. We consider the potential profile of the effective potentials 
$V\T$ and $V\S$ in a Schwarzschild black hole case for the gravitational and the 
scalar fields respectively. The Regge\,-Wheeler equations \reff{schroedTens} 
and \reff{schroedScal} can be made dimensionless by dividing through by 
the black hole mass $m$. In this way the potentials \reff{Gravitypot} and \reff{Scalpot} 
become
\bea
V\T&=&\bra{1-\frac{2}{r}}\bras{\frac{\ell\bra{\ell+1}}{r^2}
-\frac{6}{r^3}}~,\\
V\S&=&\bra{1-\frac{2}{r}}\bras{\frac{\ell\bra{\ell+1}}{r^2}
+\frac{2}{r^3}+u^2}~,
\label{normalpotential}
\eea
where we have defined (and dropped the primes),
\be
\kappa' = m\,\kappa~, \qquad r' = \frac{r}{m}~, \qquad u = m\,\ti{U}~.
\label{dimensionless}
\ee
For the gravitational perturbations and the scalar perturbations with 
$u = 0$, the derivative of the potential has two roots with one in 
the unphysical region $r<0$ and the other one in the region $r>0$ 
corresponding to a maximum of the potential. For the scalar perturbations 
with $u \ne 0$, the potential has three extrema: one in the unphysical 
region $r<0$, a local maximum at $r_{max}$ and local minimum at 
$r_{min}$ in the region $r>0$ such that $2<r_{max}<r_{min}$. 

Fig \ref{tensorpotential} shows a plot of the potential for the gravitational 
field for different $\ell$ as a function of the Schwarzschild radial coordinate 
$r$ in (a) and the tortoise coordinates $r_*$ in (b). In this case the potential 
decays exponentially near the horizon and as $1/r^2$ at spatial infinity. 
\begin{figure}[!!ht]
 \begin{minipage}[b]{6cm}
  \includegraphics[scale=0.4]{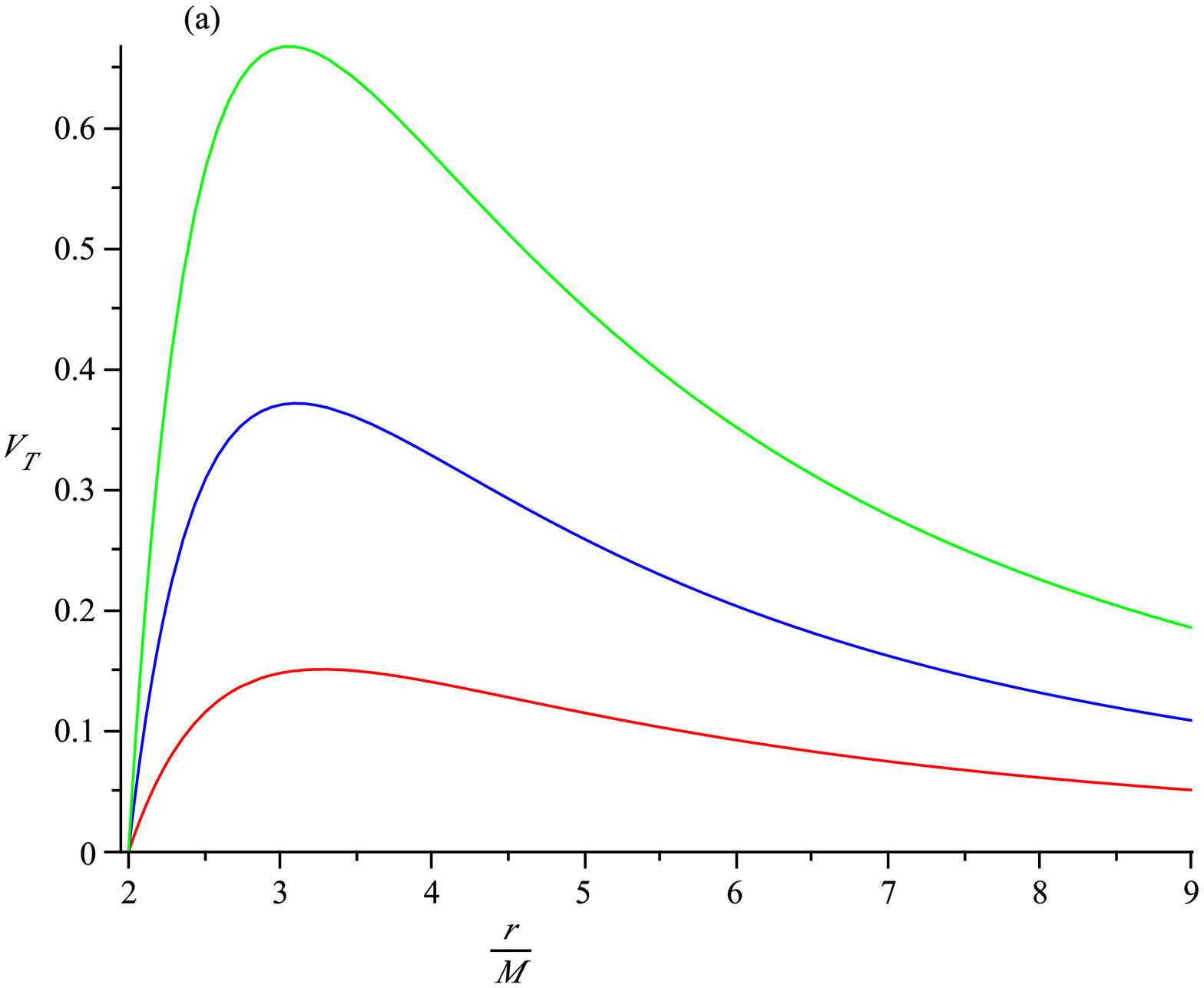}
 \end{minipage}
 \begin{minipage}[b]{8cm}
  \includegraphics[scale=0.4]{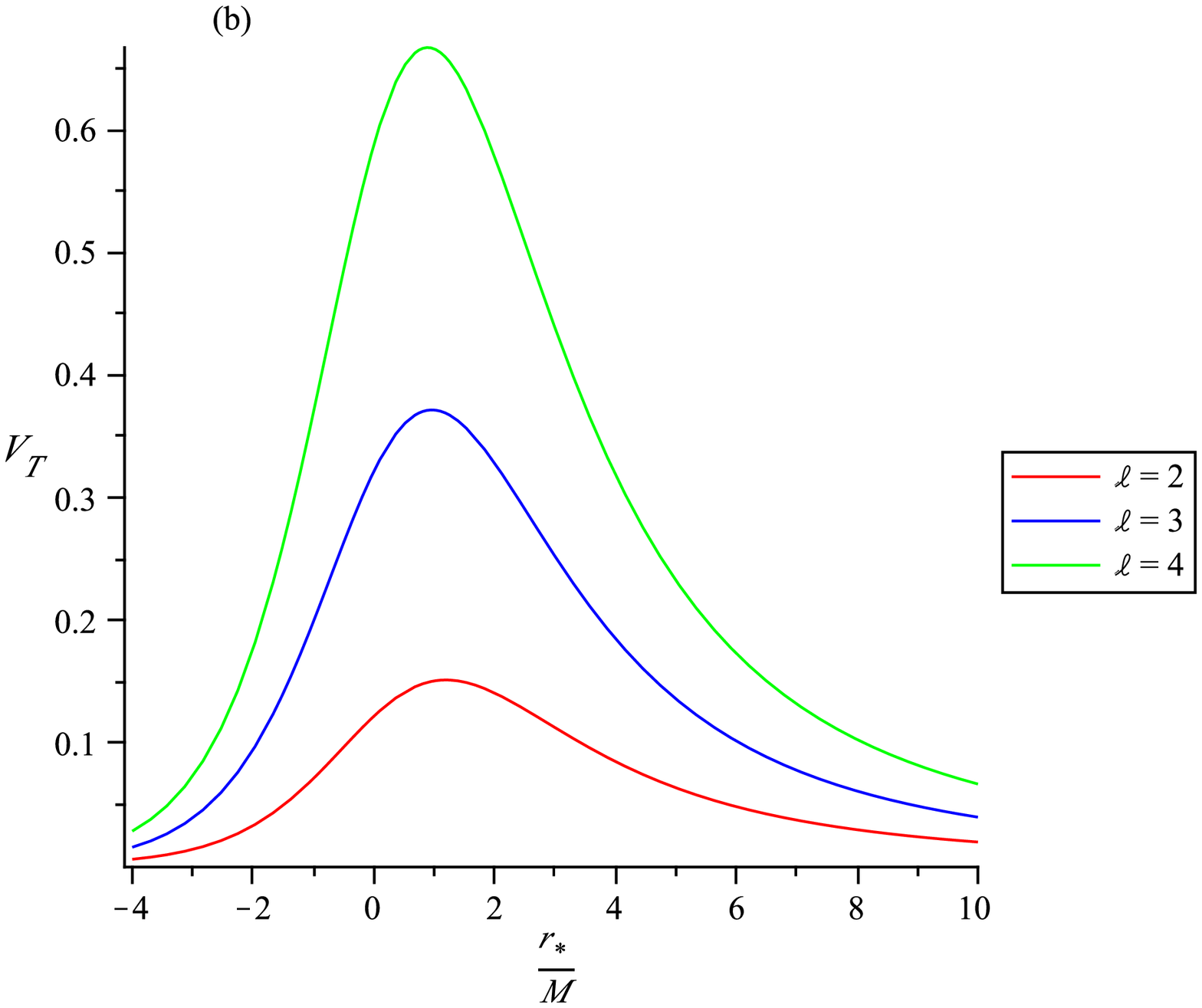}
 \end{minipage}
    \caption{\small The potential for the gravitational field for $\ell = 2, 3, 4$ 
as a function of $r$ (a) and $r_*$ (b).}
    \label{tensorpotential}
 \end{figure}
 
Fig \ref{scalarpotential} shows the potential profile for the scalar field for 
several values of $u$ at $\ell = 2$ in (a) and at $\ell = 4$ in (b). We see that 
the effect of the massive term $\ti{U}$ is to move the asymptotic value 
of the potential of scalar perturbations up by $u^2$ and to cause the 
potential to approach the asymptotic value slowly. Moreover, increasing 
the value of $u$ causes the peak of the potential to broaden as the peak 
value decreases relative to the asymptotic value. The peak eventually 
disappears altogether when $u$ exceeds a certain value.
\begin{figure}[!!ht]
 \begin{minipage}[b]{6cm}
  \includegraphics[scale=0.4]{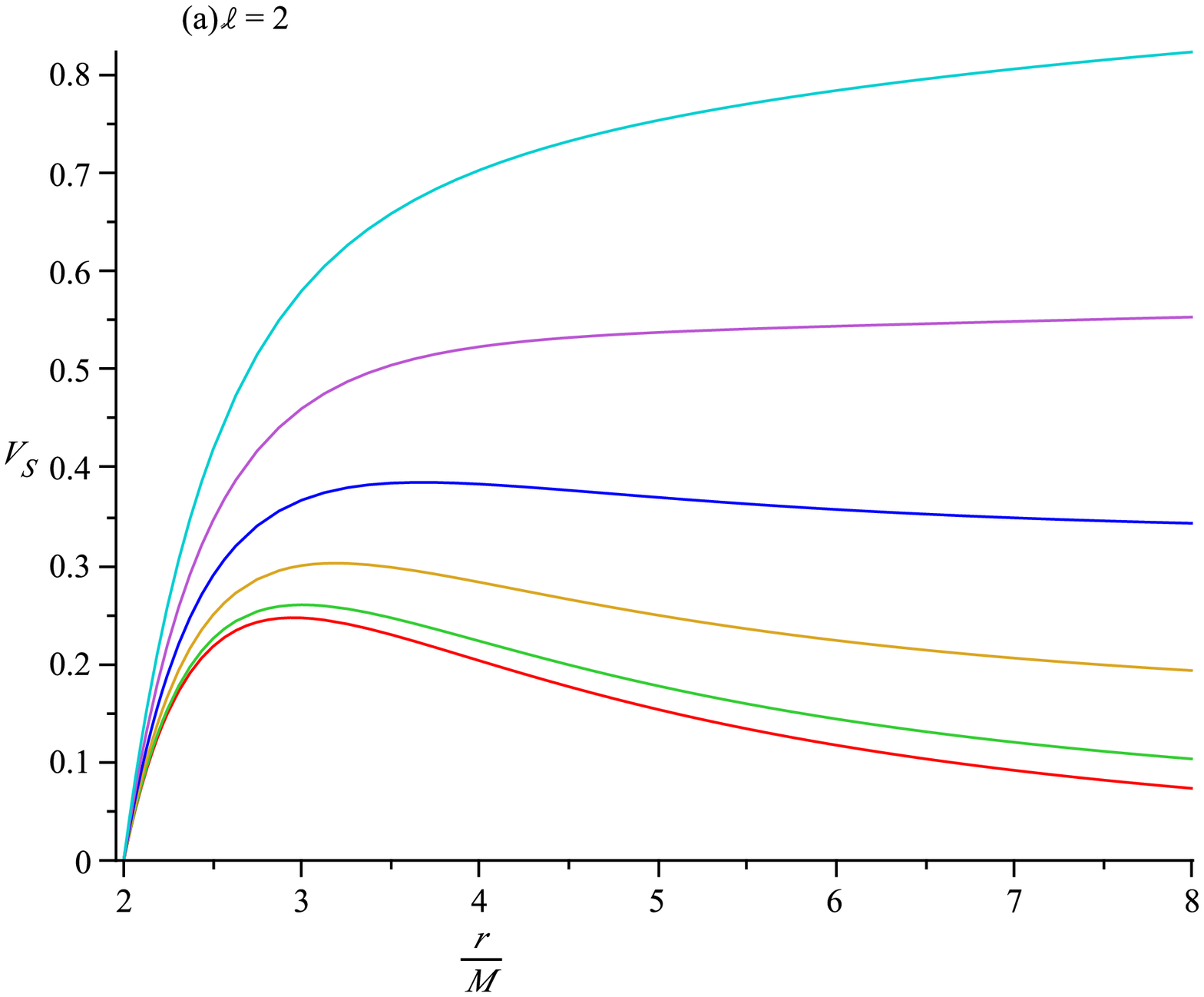}
 \end{minipage}
 \begin{minipage}[b]{8cm}
  \includegraphics[scale=0.4]{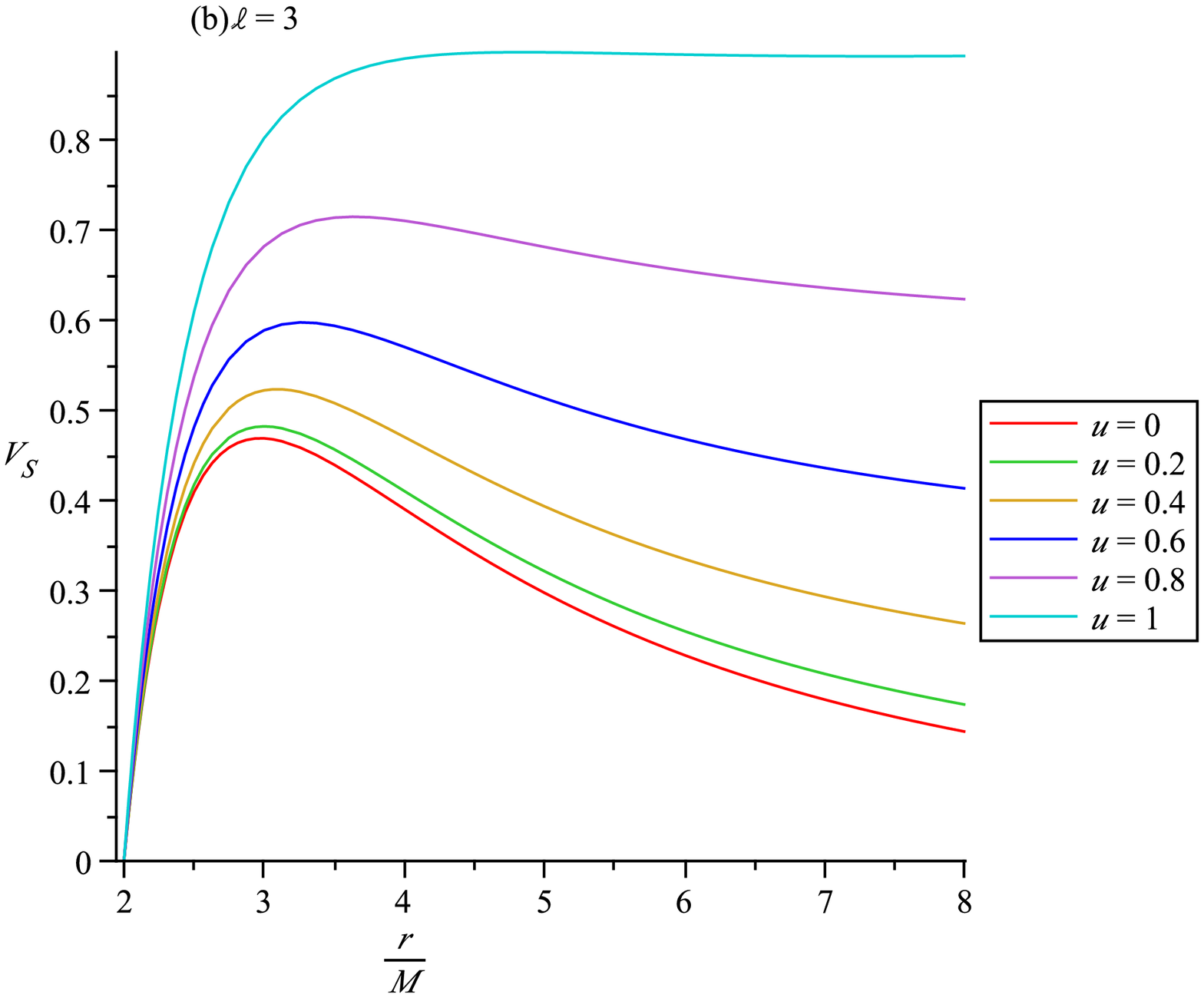}
 \end{minipage}
    \caption{\small The potential for the scalar field for different $u$ as a 
    function of $r$ for $\ell = 2$ (a) and $\ell = 3$ (b).}
    \label{scalarpotential}
\end{figure}

\subsection{Black hole stability}
\label{secStability}
We now investigate the stability of the black hole to external perturbations which 
depends on the black hole remaining bounded in time as it evolves. 
The asymptotic behaviour of the solutions to \reff{schroedTens} is 
given as
\be
M \sim e^{\pm  \,i\kappa r_* }~,
\ee
both at the horizon and at spatial infinity. If we consider purely imaginary 
solutions such that we set $\kappa = - \,i \alpha$, then the time dependence 
of the perturbations evolves like $e^{\alpha t} $, which is unstable owing to the 
fact that they grow exponentially with time.  For regularity, we require the 
perturbation to fall off to zero at spatial infinity and therefore choose 
\be
M \sim e^{- \alpha r_*}~.
\label{solinfty}
\ee
If \reff{solinfty} is to be matched to the solution that goes to zero at the horizon, 
then $\pd M/\pd r_*<0$, $\pd^2 M/\pd r_*^2<0$ within the range $-\infty$ to $\infty$. 
However, this is not the case since the potential is positive definite and as a 
result \reff{schroedTens} never becomes negative in this range. Since the 
solutions cannot be matched, this rules out perturbations that grow 
exponentially with time. This proof of stability of a black hole was first provided by 
\cite{Vishveshwara:1970}. Later on \cite{Detweiler:1973, Wald:1979} provided 
a more rigorous proof using the energy integral. This can be derived by first 
considering the time dependent version of \reff{schroedTens}
\be
\bra{\frac{\pd^2}{\pd t^2}  -\frac{\pd^2}{\pd r_*^2} +V\T\, }M= 0~.
\label{RWtime}
\ee
(recalling that the time dependence was replaced by the factor 
$e^{i \omega t}$ when we considered time harmonics). Multiplying \reff{RWtime} 
by the partial derivative of the complex conjugate $M^*$ with respect to time 
and then adding the resulting equation to its complex conjugate we get	
\be
\frac{\pd}{\pd r_*} \bra{\frac{\pd M^*}{\pd t}\frac{\pd M}{\pd r_*} 
+ \frac{\pd M^*}{\pd t}\frac{\pd M}{\pd r_*} } = 
\frac{\pd}{\pd t} \bra{\abs{ \frac{\pd M}{\pd t} }^2
+ \abs{\frac{\pd M}{\pd r_*} }^2 + V\T\, \abs{M}^2} ~.
\label{absolute}
\ee
After integration by parts over $r_*$ from $-\infty$ to $\infty$, the left-hand 
side of \reff{absolute} vanishes and we obtain the {energy integral},
\be
\int^{\infty}_{-\infty} \bra{\abs{ \frac{\pd M}{\pd t} }^2
+ \abs{\frac{\pd M}{\pd r_*} }^2 + V\T\, \abs{M}^2} dr_* = constant~.
\label{energyint}
\ee
Since $V\T$ is positive definite, the integral \reff{energyint} bounds the 
integral of $\abs{\pd M/\pd t}^2$ and it therefore excludes exponential growing 
solutions to \reff{schroedTens}. The above energy integral argument for stability 
falls short of a complete proof as it does not rule out perturbations that grow 
linearly with $t$. Also, since we have only provided the bounds for integrals of 
$M$, the perturbation may still blow up as $r \rightarrow \infty$. 
The best proof of black hole stability was provided by Kay and Wald \cite{Wald:1987} 
which,unlike the energy integral proof, proved that $\psi$ remains pointwise bounded 
when \reff{schroedTens} is evolved from a smooth, bounded initial data.\\
The proof of stability for the scalar perturbations depends on $\ti{U}$. The potential 
$V\S$ in \reff{Scalpot} remains positive definite subject to the condition
\be
\ti{U}^2= \frac{\f_1}{3\,\f_2} \geq 0~.
\label{Scalstability}
\ee
There could also be tachyonic instabilities associated with these 
modes if $\f_1 \leq 0$. Both these instabilities do not arise, however, as we have 
shown in \cite{Nzioki:2014} that the necessary conditions for the existence of a 
Schwarzschild black hole solution in $f(R)$ theories are consistent with the requirement 
that $\f_1 >0 $ and $\f_2 >0 $.
\subsection{Quasinormal modes}
\label{secQNM} 
The gravitational quasinormal modes (QNMs)  are solutions to the Regge\,-Wheeler 
equation \reff{schroedTens} subject to the boundary conditions
\begin{equation}
M\sim\begin{cases}
    ~~e^{i\kappa r_*} \quad &\mbox{for}~~~r_*\rightarrow-\infty \\
     ~~e^{-i\kappa r_*} \quad &\mbox{for}~~~r_*\rightarrow+\infty~.
  \end{cases}
  \label{TensorBC}
\end{equation} 
These boundary conditions represent purely outgoing 
waves at infinity ($r\sim r_*\rightarrow\infty$) and purely ingoing waves at the horizon 
($r\rightarrow 2m$,~$r_*\rightarrow-\infty$). In other words we want to discard unwanted 
contributions at the event horizon and at spatial infinity as we do not want gravitational 
radiation entering the spacetime from infinity to continue to perturb the black hole, nor do we 
want waves coming from the vicinity of the horizon. 

Obtaining solutions to \reff{schroedTens} and \reff{schroedScal} requires discrete values 
of the frequency parameter $\kappa$ called quasinormal frequencies belonging to the 
quasinormal modes of the black hole. The quasinormal 
frequencies have both a real and imaginary part which we write as 
\be
\kappa= \Re(\kappa) + \Im(\kappa)~.
\ee
Since QNMs are characterised by the parameters of the black hole \cite{Vishveshwara:1970}, 
we expect the imaginary part to be damped with time for each value of $r_*$ due to energy 
being radiated to infinity or the horizon. If we then consider that in \reff{schroedTens} and 
\reff{schroedScal} that the time dependence has been replaced by the factor $e^{i \omega t}$, 
we expect to have $M \sim e^{i\kappa(t - r_*)} $ at spatial infinity.  We see from 
this that $\Im(\kappa)<0$ corresponds to a bound state since the solution 
\reff{TensorBC} vanishes exponentially for $r_*\rightarrow+\infty$. This option for a negative 
imaginary part is excluded since the potential $V\T$ decays towards spatial infinity 
and therefore disallows these bound states. We can therefore only have $\Im(\kappa)>0$ 
which corresponds to the solution being damped with time but diverges exponentially as 
$r_*\rightarrow+\infty$ on a hypersurface of constant time; the same holds for the horizon. 
This consequence of divergence is balanced out by the fact that it takes the signal an infinite 
time to reach, for example, spatial infinity.

The scalar QNMs correspond to solutions of \reff{schroedScal} with 
\begin{equation}
\R\sim\begin{cases}
    ~~e^{i\chi r_*} \quad &\mbox{for}~~~r_*\rightarrow-\infty \\
     ~~e^{-i \chi r_*} \quad &\mbox{for}~~~r_*\rightarrow+\infty~,
  \end{cases}
  \label{ScalarBC}
\end{equation}
where 
$\chi = \sqrt{\kappa^2-\ti{U}^2}$ for the scalar field. For the choices $\Im(\kappa) \approx 0$ 
and $\kappa \leq \ti{U}$, there will be no energy radiating into infinity. The sign of $\chi$ is 
chosen so as to be in the same complex surface quadrant as $\kappa$.

There have been numerous attempts to calculate QNMs to high accuracy using numerical 
and semi-analytical methods. Difficulties arise from, for example, the admixture of the 
solutions such that the exponentially growing required solution gets contaminated by 
traces of the unwanted solution which decreases exponentially as we approach the 
boundaries. In 1975, Chandrasekhar and Detweiler \cite{Chandrasekhar:1975} computed 
numerically the first few modes and in 1985, Leaver \cite{Leaver:1985} proposed the most 
accurate method to date. Other methods have been employed in 
 \cite{Leaver:1985, Nollert:1993, Nollert:1992, Mashhoon:1983, Ferrari:1984, Ferrari:1984b, 
 Schutz:1985,Iyer:1987, Iyer:1987b, Froman:1992,Motl:2003}.
 Comprehensive reviews on black hole QNMs can be found in 
\cite{Nollert:1999, Kokkotas:1999, Berti:2009, Konoplya:2011}.

For the scalar field perturbations, studies have shown that the mass of the field has 
crucial influence on the damping rate of the QNMs. Using the WKB approximation 
\cite{Iyer:1989, Simone:1992,Konoplya:2002}, it was found that when the massive 
term $u$ of the scalar field increases, the damping rate decreases.
Later calculations, using the continued fraction method by Leaver \cite{Ohashi:2004, Konoplya:2005}, 
showed that as a result of the decreasing damping rates, for certain values of $u$, there are 
QNM oscillations that are `almost' purely real modes with arbitrary long life.

In GR the possible sources of massive scalar QNMs are from the collapse of 
objects made up of self-gravitating scalar fields (`boson' stars) 
\cite{Colpi:1986,Friedberg:1987,Seidel:1991}, in situations where the massless 
field gains an effective mass \cite{Konoplya:2008} or as  
 scalar field dark matter \cite{Cruz:2011}.
In order to illustrate what these results mean for $f(R)$ theories of 
gravity we restrict our attention to the $\ell = 0$ multipole of the field. From 
\cite{Ohashi:2004}, the cut-off mass at which the QNMs 
disappear for these modes is approximately at $m\,\ti{U} = 0.4-0.5$ and from PPN 
constraints \cite{Clifton:2008} for these theories we obtain the bound for $\ti{U}$ 
as  
\be
\ti{U}^2=\frac{\f_1}{3\f_2} >>\frac{2}{L^2}
\label{PPNbound}
\ee
where $L$ is the smallest length scale on which Newtonian gravity has been observed. 
Recent results \cite{Geraci:2008} place at $L \sim 10\,\mu m$ and using this we can 
set \reff{PPNbound} as
\be
\ti{U} \gg1.4 \times10^5\,m^{-1}
\ee
Given these details, we can estimate that the mass of the black hole associated with the 
disappearance of the QNMs
\be
{\mathrm{BH ~mass}} \ll 4\,\mu m~.
\ee
Such a black hole could only have been formed from density fluctuations in the early universe 
\cite{Hawking:1974,MacGibbon:1990}. 
Furthermore,  if these primordial black hole are to be detected now, they would have to have an 
initial mass of subatomic scales ($\sim 10^{-16} m$) \cite{Hawking:1971}. These results 
apply to QNMs at lower overtones and even then, QNMs are short-ranged, making their 
detection currently unfeasible \cite{Konoplya:2011}.    

\section{Solutions to the perturbation equations} 
\label{chp:Sol}
\subsection{The structure of the equations}
The structure of the system of governing equations for the perturbations is made up of covariant 
and gauge invariant evolution, propagation and
constraint equations. The true 
degrees of freedom of this system is governed by the reduced set of 
master variables $M$ and $R$, which obey the tensorial 
equations \reff{RMtensorwave} and \reff{Curvetracenl2}, respectively. All other variables are then 
related to these master variables by quadrature, plus frame degrees of freedom.
Harmonic expansion of the perturbation equations allows us, at any radial
position from the black hole, to present the equations in matrix form. The 
harmonic variables in \reff{oddvars} and \reff{evenvars} can then be treated 
as the basis of a 34-dimensional 
vector space~${\cal V}_{34}$. We can then analyse the system of equations to obtain solutions.
In this section, we present the procedure for this analysis, as set out in 
\cite{Clarkson:2003}. 
\begin{itemize}
\renewcommand{\labelitemi}{$-$}
\item{After adopting spherical harmonic decomposition, 
the number of variables in the system of equations 
is 34 in total. Let $\mathbf{V}$ denote the 34-dimensional vector 
consisting of these odd $\vo$ and even $\ve$ variables as presented in 
\reff{oddvars} and \reff{evenvars} respectively, such that 
\be
\mathbf{V}=(\mathrm{Odd~variables}~|~\mathrm{Even~variables})=(\vo,\ve)~.
\ee }
\item{We use the time harmonics in these equations 
which results in:
\begin{itemize}
\item[{\tiny{{$^\bullet$}}}] 29 propagation equations which constitute a linear 
system of ODEs
\be
\hat{\mathbf{V}}_{29} = \mathbf{P \, V }~,
\label{propgeneral}
\ee
where $\mathbf{V}_{29}$ is a vector consisting of the 29 elements
of~$\mathbf{V}$ which have a propagation equation and 
$\mathbf{P}$ is a $29\times34$ propagation matrix in which the evolution 
equations, where the dot derivatives are replaced by $i\omega$, contain 
hat derivatives in them.
\item[{\tiny{{$^\bullet$}}}] 25 algebraic relations between the variables, made up of 
18 evolution equations as well as 7 constraints. These, in matrix notation, take the form
\be
\mathbf{F\,V}=\mathbf{0}~,
\label{Eeq}
\ee
where $\mathbf{F}$ is a $25\times34$ matrix. Since the 
constraints propagate and evolve consistently, this means that the rows 
that make up the constraints are really linear 
combinations of the 18 rows that make up the algebraic relations derived 
from the evolution equations (this excludes the constraint \reff{Zconst} 
since there is no evolution equation for $Z^a$). As a result, 6 
of the rows in $\mathbf{F}$ give no additional information, resulting 
in $\mathbf{F}$ being of rank 19.
 \end{itemize}
\item{So far, the formulation has resulted in 34 unknowns and 19 algebraic relations 
in the system which corresponds to $34-19=15$ degrees of freedom. This means 
that there are 15 variables that need to be solved for, which we denote by $\mathbf{v}$, and write
\be
\mathbf{V}=\mathbf{C\,v}~,
\label{solform1}
\ee
where $\mathbf{C}$ is a $34\times15$ matrix of the form
\be
\bra{
\begin{array}{c|c}
  \begin{array}{lcr}
   \longleftarrow & 6 & \longrightarrow
  \end{array} &
  \begin{array}{lcr}
    \longleftarrow~ & ~9~ & ~\longrightarrow
  \end{array} \\
  & \\
  \mbox{\sc odd} & \mbox{\sc even}\\
  &
\end{array}
}.
\ee
}
\item{
We now split the vector $\mathbf{v} $ into two parts: 
$\mathbf{v} = (\mathbf{v}_D,\mathbf{v}_F)$, the first one $\mathbf{v}_D$ containing the 
10 variables which have an individual propagation equation and the second one
$\mathbf{v}_F$ the $15-10=5$ variables that do not. The latter part corresponds to 
5 frame degrees of freedom. Inserting
\reff{solform1} into the propagation equation, \reff{propgeneral} yields 
the underlying propagation equation for the solution vector as
\be
\hat{\mathbf{v}}_D=\mathbf{B}\,\mathbf{v}_D+\mathbf{A}\,\mathbf{v}_F,
\label{solform2}
\ee
where $\mathbf{B}$ is a $10\times10$ matrix and $\mathbf{A}$  is 
$10\times5$. }
\item Finally, since we have the freedom to choose the 5 frame basis 
$(\mathbf{v}_F)$, we find that there are only $10-5=5$ true dynamical
propagation equations to solve for the unknown 5 components of $\mathbf{v}_D$}
\end{itemize}

\subsection{Determining the full solution}
\label{secSol}

\subsubsection{Odd}
\label{oddsolssec}
The problem of finding a solution lies in deciding which variables to choose as the basis. 
To concur with \cite{Clarkson:2003} for the GR case, we will choose the frame in which $\bar
Y\V=\bar\A\V=0$ and as a result $\xi\S=\Omega\S=\bar
a\V=\xbar{W}\V=\bar Z\V=\Omega\V=0$.
The basis vector for the solution is chosen to be
\be
\mathbf{v}= \bra{
\begin{array}{c}
  \xbar M\T \\
  \hat{\xbar M}\T
\end{array}
};
\ee
According to \reff{solform1}, the remaining variables in terms of this 
solution basis vector are given by 
\be
\bra{\begin{array}{c}
 \bar\E\T \\ \H\T \\ \bar\Sigma\T \\ \bar\zeta\T\\
 \bar\E\V \\ \H\V \\ \bar\Sigma\V \\ \Omega\V\\ \bar\A\V \\ \bar\alpha\V\\
 \bar\hatn\V 
 \\ \xbar{W}\V \\ \bar Y\V \\ \bar Z\V
 \\ \H\S \\ \Omega\S \\ \xi\S
\end{array}}=\bra{\begin{array}{cc}
    -J/2\phi^2 r^4   &  -2/\phi r^2     \\
   \bra{-4L+J+8r^2\omega^2+16}/4{i\omega\phi r^4} ~    &   ~
   -J/{2i\omega\phi^2r^4}    \\
    1/i\omega r^2    &   2/i\omega\phi r^2     \\
      2/\phi r^2   &   0     \\
      \l/\phi r^3   &  0      \\
     0& -\l/i\omega\phi r^3   \\
      -\l/i\omega\phi r^3    &0      \\
      0&0     \\
      0 & 0\\
       \l/i\omega\phi r^3 & 0 \\
      0   &     0   \\
      0   &    0    \\
      0   &    0    \\
      0   &    0    \\
      - L \l/i\omega\phi r^4   &    0    \\
      0   &   0     \\
      0   &0
\end{array}}\bra{\begin{array}{c}
  \xbar M\T \\
  \hat{\xbar M}\T
\end{array}}
\ee
where for the sake of brevity we have used the aliases
\bea
J&=& 3\phi^2\,r^2-4~,\\
L &=& \ell\bra{\ell+1}~,\\
\l &=& \bra{\ell-1}\bra{\ell+2}= L-2~.\\
\eea

\subsection{Even}
\label{evensolssec}
As in the odd case, we choose the frame $\A\V=Y\V=0$ (and hence 
$Z\V=0$). We will choose  
\be
\mathbf{v}= \bra{\begin{array}{c}
   M\T \\ 
   \hat{M}\T\\
  R\S\\
  \hat{R}\S
\end{array}}~,
\ee
as the basis vector for the full solution. 
The expressions for the obtained solutions are large and so in the interest 
of brevity we introduce the variable $\MM$ as a function of the basis variables such that
\bea
\MM&=&\frac{ 1}{24 c_3 \f_1\bra{ L^2\, \l^2- \A^2 ( 4 L+4 - c_3)^2 r^4 \,\omega^2}}   
\brac{  -i \, \omega \, \phi \,r [96\, L \,\l  (L+1) - 3 (8 \l \,(L+4)  \right. \nn\\
&& \left. +3  (8 L-16  - c_3)c_3) \phi^2 r^2] \f_1\, M\T 
-72\, i  \omega \, \A \, \phi^3 \,r^5 \, c_3\, \f_1\, \hat{M}\T 
\right. \nn\\
&& \left.  - i \, \omega\,\phi \,r [(8 \,\l(L+4) 
+ (8 L-16  - c_3) c_3) \phi^2 \,r^2-32  L\, \l (L+1) ]  \f_2 \, R\S \right. \nn\\
&& \left. + \,(24\, i \,\omega \,\A \, \phi^3 r^5 ) c_3 \,\f_2 \,\hat{R}\S}~.
\label{newM}
\eea
with the solution given by:\\

{\small
\renewcommand{\arraystretch}{1.2}
\noindent
\resizebox{\linewidth}{!}{
  \begin{tabular}{c | c | c| c| cl } 
\mbox{\sc EVEN}   &  ${\MM}$  &  $ M\T$   &   $ R\S$   &    $X\S$    \\ \hline
 $\E\T$	& $-\frac{3  L \l J+ (J-8) (J+4) \omega^2 r^2 }{6 i \omega \phi^3 r^5}$
  		&$\frac{(4 \l c_3 -32 \l( L+1)  + c_3^2) }{2 ( 8 L +8+ c_3) \phi^2 r^4} $
 		&  $-\frac{\Delta }{18 L \l r^2 c^2_3 \f_1}$
 		& $\frac{(4 L -8 - c_3)  ( 3 r-2 )\f_2}{9  \phi r^3 c_3 \f_1}$    
		\\ 
$ \bar\H\T$  	& $-\frac{(J - 4 L) (J + 4 \l) }{8 \phi^2 r^3 }$   
			&$-\frac{2 i \omega}{\phi^2 r^2} $
			&   0
			& 0 
			\\ 
$ \Sigma\T$   	& $- \frac{2  L \l+ (J-8) \omega^2 r^2}{2\omega^2 \phi r^3 } $  
			&$\frac{4 \l }{i \omega c_3 r^2 }$
			&    $-\frac{ 2 [3 (\l - c_3) \f_2- r ( 3 r-2 ) (\f_1 + 6 \f_2 \omega^2)]}{9 i \omega  r^2 c_3 \f_1} $
			& $\frac{2  (J-8)\f_2}{ 9 i  \omega \phi r^3 c_3 \f_1 }$ 
			\\  
  $\zeta\T $ 	&$ \frac{ 2 L \l}{i \omega r^3 \phi^2 }$
  			&$\frac{2}{\phi r^2} $
			& $\frac{  2 (4 L +4- c_3) \f_2}{3 \phi r^2 c_3 \f_1} $
			&$-\frac{4 \f_2}{c_3 \f_1}$  
			\\ 
$\E\V$	&$ -\frac{  L  \l^2 }{i \omega \phi^2 r^4 }$
		&$\frac{ \l}{\phi r^3 }$
		&$\frac{  (4 L  +4-c_3 ) (c_3- 4 L +8)\f_2 }{12\phi r^3  c_3 \f_1}$ 
		&$\frac{( 4 L-8  - c_3) \f_2}{2  r c_3 \f_1} $  
		\\ 
$\bar\H\V$	& $\frac{\l (J-8) (L + 2 \omega^2 r^2 )}{8 \omega^2 \phi r^4 }$   
			 &$-\frac{\l(  J-8)  }{2 i \omega r^3 (4+c_3)  }$
			&   $\frac{( J-8) [3 (\l- c_3) \f_2 -r (3 r-2) (\f_1 + 6  \omega^2 \f_2)]}{36 i \omega  r^3 c^2_3 \f_1 }$
			& $ \frac{ ( J-8)   \A \f_2  }{3 i \omega r^2 c_3 \f_1  }$
			\\
 $ \Sigma\V$	&$-\frac{ L \l [(2 L-4 ) c_3+ ( J-8) \E r^2]}{2 \omega^2 \phi^2 r^4 c_3 }$
 			& $\frac{\l}{i \omega \phi r^3  }$
    			&$\frac{\Pi }{6  i \omega \phi r^3 c^2_3 \f_1 } $
    			&$\frac{ [  c_3 (8 \A^2 + (2 L-4  - c_3) (4 \E - \phi^2)) -16 \l \E  ] \f_2 }{i \omega r (4 \E - \phi^2) c_3^2 \f_1 }$
			\\  
$ \bar\Omega\V$	& $  - \frac{L\l  (J-8) \A }{2c_3  \omega^2 r^2 \phi }$
				&$0$
     				&$  \frac{ \A (2 c_3 r^2 \f_1  + [8 \l( L+1) + (( L+10) c_3 - 6 c_3 \omega^2 \phi^2) ] \f_2 )}{i \omega 3 r c_3^2 \f_1} $
    			 	&$  \frac{\A[(J-8) c_3-8 \l( J+4)  ] \f_2}{6 i \omega \phi r c_3^2 \f_1 }   $
     				\\ 
$\A\V$ 	&  0   &   0	  & 0   &0
		\\
$\alpha\V$	&  $  -\frac{ L \l ((4 L - c_3))}{4 \omega^2\phi^2 r^4 }  $
			&$\frac{\l(4 L - c_3) }{i \omega \phi r^3  c_3 }$
     			&$\frac{ \Psi }{ 36 i \omega \phi r^3 c^2_3 \f_1  }$
     			&$- \frac{  [\A (4L- c_3) (4 \A + \phi) -24 \omega^2 r ]\f_2}{6 i \omega  c_3 \f_1}$
     			\\
$\hatn\V$		&$ -\frac{ L\l  (J-8) }{i \omega c_3 r^2 \phi }  $
			&0 
			&$-\frac{ 4c_3r^2  \f_1 +2 (8 \l(1 + L) + ( L+10) c_3 - 6 c_3 \phi^2 \omega^2)  \f_2 }{3 r c_3^2 \f_1 } $
			&$\frac{[8 \l (J+4) -c_3 (J-8) ] \f_2 }{3  r\phi c_3^2 \f_1}$
			\\ 
$ W\V$ 	&$  \frac{ L\l  (J-8) }{4i \omega r^4 \phi }$
		 &    0 
		& $ \frac{ (J-8)(L +1) \f_2}{3 r^3 c_3\f_1}  $
		&$-\frac{ (J-8)\phi  \f_2}{2 r c_3\f_1} $  
		\\
  $Y\V$ 	&  0   &   0& 0 &0
  \\
  $ Z\V$	&  0   &   0& 0 &0
  \\
  $  \Sigma\S$	&$ \frac{ L\l(J-8)[3 c_3 -8 (\A^2-\E)r^2] }{12 \omega^2 \phi r^3 c_3} $ 
  			&0 
     			&$   \frac{\chi }{ i  \omega r^2 c^2_3 \f_1 }    $
      			&$-\frac{ [(32 \l(J+4) -4 ( J-8) c_3) (\A^2 - \E) + 9 \phi^2 (8 - J + c_3) c_3]\f_2 }{i \omega18 c_3^2 \f_1 \phi }$
			\\ 
$ \theta\S$ 	& $    -\frac{ L\l \ell (J-8) (\A^2-\E)}{\omega^2 r \phi c_3 }$
			& 0
      			&$\frac{\Gamma}{ i \omega r c^2_3  \f_1 }$
       			&$\frac{4 \E (12 \A^2  r^3 c_3+ [6\l (J+8) - ( J-8) c_3] r  -3 c_3^2 )\f_2 }{i \omega (4 + J)  (4 \A + \phi) r c_3^2 \f_1 } $
			\\
  \end{tabular}%
  }}
  
  where
\bea
\Gamma&=& c_3^2 \f_1 + 4 (\A^2 - \E) [ (2\l ( J+4) + 3 (L+2) c_3) \f_2 + \frac{  8 c_3 \f_1}{ \phi^2 -4 \E }] r \nn\\
&&+\, \frac{2  [(4 + J) \E + 3 c_3 r] \omega^2 c_3 \f_2}{r}~,\\
\chi &=& 1/72 \{36  ( L (c_3+8)+8 - c_3 )c_3 \f_2 + 64 (\A^2 - \E) r^2 (\l( J+4) \f_2 + c_3 \f_1 r^2) \nn\\
 &&-\, 3  [( J+4) (8 L+8 - c_3) \f_2 - 4  (c_3 \f_1 - \f_2 \E (16 + 8 L + 3 c_3 - 16  \omega^2\phi^2) \nn\\
 &&+\, 8 \A^2 \f_2 (2 + \l \omega^2\phi^2) )  r^2  ]c_3  \}~,\\
 \Pi &=& (2 \l  - c_3) (4 L+4  - c_3) c_3 \f_2 + 2 \E [2 c_3 \f_1 r^4+ (2 \l (4 + J) + 3 (2 + L) c_3) r^2\f_2  \nn\\
 &&+\, 2 c_3 (-4 - 4 L + c_3) \f_2 \omega^2]~,\\
\Psi &=& 3 (4 L - c_3) [4 (J - 4 L) (L+1) + (J + 2 L) c_3 - c_3^2] \f_2 - 2 c_3 r ( 3 r)-2 [(4 L - c_3) \f_1 \nn\\
 &&+\, 2 (8 - 4 L + c_3) \f_2 \omega^2]~,\\
\Delta &=&3 r^2 \{4  L \l c_3 \f_1 + 6 L\l(-6 - 6 L + c_3)  \phi^2 \f_2+ [16 (L+1)^3 + 8 (2 (L-4) L -1 ) c_3 \nn\\
 &&+\, (L+1) c_3^2]\omega^2 \f_2 \} + 6  L \l(4 + 4 L - c_3) (6 + 6 L - c_3) \f_2 - 8 l L c_3 r (\f_1 + 6 \f_2 \omega^2) ~.\nn\\
\eea

\section{Discussion}
\label{chp:Con}
We used the 1+1+2 covariant approach to GR to give a detailed analysis of linear perturbations 
of Schwarzschild black holes in $f(R)$ gravity. 

Since the background only involves scalar quantities,  all vector and tensor quantities 
are gauge invariant under linear perturbations (as a consequence of the Stewart and Walker Lemma). 

We were able to obtain a frame invariant TT tensor $M\T$ which satisfies  the Regge-Wheeler equation 
irrespective of parity and demonstrated that for the tensor modes, the underlying dynamics in $f(R)$ gravity is governed 
by a modified Regge-Wheeler tensor which obeys the same Regge-Wheeler equation as 
in GR. In order to close the system a scalar wave equation for the Ricci scalar must be included which corresponds 
to the propagation of the additional scalar degree of freedom not present in GR. 
Since the Regge-Wheeler equation governs the odd (axial) perturbations. 

The main difference between GR and $f(R)$ gravity is the appearance 
of scalar perturbations representing the propagation of the additional gravitational degree of 
freedom not present in GR. This extra mode introduces a ghost problem but this can be avoided 
in $f(R)$ theories of gravity if the condition $\partial f/ \partial R >0 $
is satisfied. Since the necessary conditions for the existence of a 
Schwarzschild black hole solution in $f(R)$ theories are consistent with the requirement 
that $\partial f/ \partial R >0  $ and $\partial^2 f/ \partial R^2 >0  $, the no-ghost condition is satisfied. 
This is in agreement with \cite{DeFelice:2011,Myung:2011} where the stability of the Schwarzschild 
solution in $f(R)$ was studied.

For the (QNMs) that follow from the scalar perturbations, we find 
that possible sources of scalar QNMs for the lower multipoles are from primordial Black Holes. 
Higher mass, stellar black holes are associated with extremely high multipoles, which can only 
be produced in the first stage 
of black hole formation. Since the scalar QNMs are short ranged, this scenario makes their 
detection beyond the range of current experiments.

Finally, we find the solutions to the perturbation equations by introducing harmonics to the 
system of linearised equations. The harmonic decomposition reduced the system into a linear 
system of algebraic equations which simplified things and we were able 
to find the solution of the system using matrix methods, while employing the freedom to choice 
of frame vectors.

\begin{center}
{\bf Acknowledgements}
\end{center}
AMN would like to thank Chris Clarkson for helpful discussions. RG acknowledges 
the National Research Foundation (South Africa) and the University of KwaZulu-Natal for financial support.

\bibliography{RW_VI}
\bibliographystyle{Science}
\end{document}